\documentclass[prd,showpacs,showkeys,nofootinbib,twocolumn]{revtex4}

\usepackage{graphicx} 
\usepackage{subfigure}
\usepackage{amsmath} 
\usepackage{amssymb} 

\DeclareMathOperator{\Order}{\mathcal{O}}

\begin{document}

\title{Influence of internal structure on the motion of test bodies in extreme mass ratio situations} 

\author{Jan Steinhoff}
\email{jan.steinhoff@ist.utl.pt}
\homepage{http://jan-steinhoff.de/physics}
\affiliation{CENTRA, Instituto Superior T\'ecnico, Avenida Rovisco Pais 1, 1049-001 Lisboa, Portugal}

\author{Dirk Puetzfeld}
\email{dirk.puetzfeld@zarm.uni-bremen.de}
\homepage{http://puetzfeld.org}
\affiliation{ZARM, University of Bremen, Am Fallturm, 28359 Bremen, Germany}

\date{ \today}

\begin{abstract}
We investigate the motion of test bodies with internal structure in General Relativity. We utilize a multipolar approximation scheme along the lines of Mathisson-Papapetrou-Dixon including the quadrupolar order. The motion of pole-dipole and quadrupole test bodies is studied in the context of the Kerr geometry. For an explicit quadrupole model, which includes spin and tidal interactions, the motion in the equatorial plane is characterized by an effective potential and by the binding energy. We compare our findings to recent results for the conservative part of the self-force of bodies in extreme mass ratio situations. Possible implications for gravitational wave physics are outlined.
\end{abstract}

\pacs{04.25.-g; 04.20.-q; 04.20.Fy}
\keywords{Approximation methods; Equations of motion; Extreme mass ratios; Variational principles}

\maketitle


\section{Introduction}\label{introduction_sec}

Approximate analytic treatments of the binary problem in General Relativity are possible in certain limited regimes. These include the weak field (post-Minkowskian), weak-field and slow motion (post-Newtonian), and extreme mass ratio approximations. An important astrophysical example of the latter  approximation is a star inspiraling into a very massive black hole, i.e., an extreme mass ratio inspiral. The zeroth order approximation in the extreme mass ratio case describes the motion of the lighter mass by a geodesic in the fixed background spacetime generated by the heavier mass. An important first correction derives from a perturbation of the background due to the small mass object, leading to so-called self-force corrections to the geodesic motion within the background metric. Recently this self-force program made progress, for example evolutions over thousands of orbits for a Schwarzschild background succeeded \cite{Warburton:Akay:Barack:Gair:Sago:2011}, see also \cite{Pound:Poisson:2007, Barack:Sago:2010}. Also self-force corrections to the innermost stable circular orbit were derived \cite{Barack:Sago:2009}, whose frequency is an important observable of gravitational wave astronomy.

Besides self-force corrections to the geodesic motion within a background metric, corrections due to the small object's spin and higher multipoles -- encoding its finite size -- arise. Multipolar corrections to the motion of test bodies were first derived by Mathisson \cite{Mathisson:1937, Mathisson:2010} and Papapetrou \cite{Papapetrou:1951:3}, and since then have been studied in the context of different approximation schemes by a number of authors \cite{Tulczyjew:1959,Tulczyjew:1962,Taub:1964,Dixon:1964,Taub:1965,Madore:1969,Dixon:1970:1,Dixon:1970:2,Dixon:1973:1,Dixon:1974:1,Dixon:1979}. It is important to systematically study these corrections in astrophysically realistic situations, in particular as the self-force calculations are approaching the second order \cite{Rosenthal:2006, Pound:2012, Gralla:2012}. In the present paper we study a truncation of the equations of motion at the quadrupole level, to which an independent derivation was given in our previous publication \cite{Steinhoff:Puetzfeld:2009:1} with the help of the multipolar approximation scheme by Tulczyjew \cite{Tulczyjew:1959}. Our analysis in \cite{Steinhoff:Puetzfeld:2009:1} made clear that there is a considerable freedom to close the equations of motion at the quadrupole level. Besides imposing a supplementary condition on the spin, corresponding to a choice for the center of the object, one must devise a model for the quadrupole dynamics.

In contrast to previous works \cite{Ehlers:Rudolph:1977,Harte:2007,Bini:etal:2008:1,Bini:etal:2008:2}, in the present paper we focus on astrophysically realistic models for the quadrupole, which derive from recent work on effective actions for extended objects. This not only includes quadratic-in-spin corrections \cite{Porto:Rothstein:2008:2,Porto:Rothstein:2008:2:err}, but also tidal interactions \cite{Goldberger:Rothstein:2006:2,Damour:Nagar:2009}, see also \cite{Bini:Damour:Faye:2012}. In fact, the interaction of a black hole's tidal field with the quadrupole of an inspiralling star may become very strong, as is indicated by the appearance of tidal disruptions. Quadrupole tidal interactions can be of more importance for gravitational wave astronomy in small mass ratio situations than spin effects -- for the measurability of spin effects see \cite{Huerta:Gair:2011}.

The present paper aims at extending the effective potential for test bodies with spin in a Kerr background \cite{Rasband:1973} to include corrections from the mentioned quadrupole models (see also \cite{Tod:Felice:1976,Suzuki:Maeda:1998} for alternative derivations, \cite{Hojman:Hojman:1977} for a charged spinning test body in Kerr-Newman spacetime, and \cite{Kunzle:1972,Barausse:Racine:Buonanno:2009} for Hamiltonian approaches). To carry out this program, it is necessary to identify various conserved quantities of quadrupolar test bodies. This is not straightforward and some of the emerging difficulties were already pointed out in our previous work \cite{Steinhoff:Puetzfeld:2009:1}. However, with an effective action as a basis one can immediately construct conserved mass and spin length parameters. Conserved quantities -- associated with symmetries (Killing vectors) of the background -- were already found in \cite{Ehlers:Rudolph:1977} and are actually the same as in the spinning test body (or pole-dipole) case \cite{Dixon:1970:1}. In a Kerr background this gives rise to a conserved binding energy and a total angular momentum of the test body.

For aligned spins and circular orbits in the equatorial plane we compute the gauge-invariant relation between binding energy and total angular momentum. This complements recent results on this relation from full numerical simulations for comparable masses \cite{Damour:Nagar:Pollney:Reisswig:2011} and from the conservative part of the self-force for extreme mass ratios \cite{LeTiec:Barausse:Buonanno:2012}. While both of these results are valid for non-spinning binaries only, we include spin and quadrupole effects (including tidal deformations) due to the smaller object in the extreme mass ratio case here. Furthermore, we compare our result with various post-Newtonian Hamiltonians or potentials. 

The structure of the paper is as follows. In section \ref{recapitulation_sec}, we briefly recapitulate some of our findings from \cite{Steinhoff:Puetzfeld:2009:1}. We then rewrite the equations of motion for a generalized supplementary condition. Subsequently, in section \ref{orbits_sec}, we work out the effective potential for spinning test bodies on equatorial orbits with aligned spin in a Kerr background. In section \ref{quadrupole_model_sec} we introduce our explicit quadrupole model. In particular we identify a mass-like quantity at this multipole order which is conserved in an approximative sense. Furthermore, we work out the effective potential in terms of a set of dimensionless parameters, and discuss the values of these parameters for astrophysically relevant situations. This is followed by a discussion of the binding energy in section \ref{sec_binding_energy}, in which we also compare our results to corresponding ones in a self-force and post-Newtonian context. We draw our final conclusion in section \ref{conclusions_sec}. In Appendix \ref{sec_misprints} we collect some misprints which we found in the literature regarding the effective potential of pole-dipole test bodies. Appendices \ref{sec_orientation}--\ref{sec_ca_results} contain supplementary material regarding our calculations. In Appendix \ref{dimension_acronyms_app}, we summarize our notation and conventions and provide a brief overview of different quantities and units used throughout the work.

\section{Basic equations}\label{recapitulation_sec}

\subsection{Equations of motion}
The equations of motion of an extended test body up to the quadrupolar order are given, see \cite{Steinhoff:Puetzfeld:2009:1} for a derivation in the context of Tulczyjew's multipolar approximation method, by the following set of equations:
\begin{eqnarray}
\frac{\delta p_a }{d s} &=& \frac{1}{2} R_{abcd} u^{b} S^{cd} + \frac{1}{6} \nabla_a R_{bcde} J^{bcde},  \label{pap1}\\
\frac{\delta S^{ab}}{d s} &=& 2 p^{[a} u^{b]} - \frac{4}{3} R^{[a}{}_{cde} J^{b]cde}. \label{pap2}
\end{eqnarray}
Here $u^a:=dY^a/ds$ denotes the 4-velocity of the body along its world line (normalized to $u^a u_a=1$), $p^a$ the momentum, $S^{ab} = - S^{ba}$ the spin, and $J^{abcd}$ the quadrupole moment with the following symmetries:
\begin{eqnarray}
J^{abcd} = J^{[ab][cd]} = J^{cdab}, \\
J^{[abc]d} = 0 \quad \Leftrightarrow \quad	J^{abcd} + J^{bcad} + J^{cabd} = 0.
\end{eqnarray}
Thus, $J^{abcd}$ has the same (algebraic) symmetries as the Riemann tensor. The corresponding stress-energy tensor of the test body can be written in the following (singular) form:
\begin{eqnarray}
&&\sqrt{-g} T^{ab} = \int d s \bigg[u^{(a} p^{b)} \delta_{(4)}- \frac{1}{3} R_{cde}{}^{(a} J^{b)edc} \delta_{(4)}	\nonumber \\
&&- \nabla_c ( S^{c(a} u^{b)} \delta_{(4)} )- \frac{2}{3} \nabla_d \nabla_c ( J^{d(ab)c} \delta_{(4)} ) \bigg]. \label{em_tensor_singular}
\end{eqnarray}
From equation (\ref{pap2}) it follows that the momentum is given by:
\begin{equation}
p^a = m u^a + \frac{\delta S^{ab}}{d s} u_b + \frac{4}{3} u_{b} R^{[a}{}_{cde} J^{b]cde}, \label{generalized_momentum}
\end{equation}
where we used $m := p_a u^a$.

\subsection{Conserved quantities}

We encountered already in \cite{Steinhoff:Puetzfeld:2009:1}, that for the pole-dipole case ($J^{abcd} = 0$), the quantity
\begin{equation}
E_{\xi} = p_a \xi^a + \frac{1}{2} S^{ab} \nabla_a \xi_b ,
\end{equation}
is conserved if $\xi^a$ is a Killing-vector, $\nabla_{(b} \xi_{a)} = 0$. In \cite[p.210]{Ehlers:Rudolph:1977} it was further shown that this is a conserved quantity even at all higher multipole orders. 

Other conserved quantities depend on the supplementary condition. For the spin length $S$ given by $2 S^2 := S_{ab}S^{ab}$ one obtains:
\begin{eqnarray}
S \frac{d S}{d s} &=& \frac{1}{2} \frac{d S^2}{d s} = \frac{1}{2} S_{ab} \frac{\delta S^{ab}}{d s} \nonumber \\
&=& S_{ab} p^a u^b - \frac{2}{3} S_{ab} R^{a}{}_{cde} J^{bcde}. \label{dSds}
\end{eqnarray}
It is easy to see that in the pole-dipole case the spin length $S$ is conserved for the two well-known supplementary conditions of Tulczyjew ($p_a S^{ab} = 0$), and Frenkel ($u_a S^{ab} = 0$).

For the mass $\underline{m}$, defined by $\underline{m}^2 := p^a p_a$, the following relations hold:
\begin{eqnarray}
\underline{m}^2 &=& m^2 + \frac{\delta S^{ab}}{d s} p_a u_b + \frac{4}{3} p_a u_b R^{[a}{}_{cde} J^{b]cde}, \label{mass_relation} \\
\frac{d \underline{m}}{d s} &=&	\frac{\delta p_a}{d s} \frac{p_b}{m \underline{m}} \left[ \frac{\delta S^{ab}}{d s}	+ \frac{4}{3} R^{[a}{}_{cde} J^{b]cde} \right] \nonumber \\
&&+ \frac{\underline{m}}{6 m} \frac{\delta R_{bcde}}{d s} J^{bcde}. \label{dmbdt}
\end{eqnarray}
The last relation follows from an insertion of (\ref{pap2}) into the expression $\frac{\delta p_a}{d s} p_b \frac{\delta S^{ab}}{d s}$ and use of $\frac{\delta p_a}{d s} p^a = \underline{m} \frac{\delta \underline{m}}{d s}$. Hence in the pole-dipole case the mass $\underline{m}$ is conserved if one chooses Tulczyjew's spin supplementary condition. However, for the Frenkel condition the mass $m$ is conserved in the pole-dipole case. 

An extension of conserved spin length and masses to the quadrupole case is one of the main obstacles in the present work. We will come back to this in sec. \ref{quadrupole_model_sec} in the context of an explicit quadrupole model.

\subsection{Spin supplementary condition and equations of motion}

A spin supplementary condition (SSC) has to be imposed to close the system of equations (\ref{pap1})--(\ref{pap2}) even in the pole-dipole approximation. Here we rewrite the equations of motion with the help of the following supplementary condition:
\begin{equation}
	S^{ab} f_b = 0. \quad \quad (\ast)\label{general_supp}
\end{equation}
This condition allows for an easy transition between the widely used Frenkel ($f_a = u_a$) and Tulczyjew ($f_a = p_a $) conditions.

Taking the total derivative of (\ref{general_supp}) leads to a relation between $p^a$ and $u^a$:
\begin{equation}
u^a = \frac{1}{p^f f_f} \left[ u^b f_b p^a + S^{ab} \frac{\delta f_b}{d s}- \frac{4}{3} f_b R^{[a}{}_{cde} J^{b]cde} \right]. \label{uprel}
\end{equation}
We now have the option to eliminate $u^a$ or $p^a$ from the equations of motion.

\paragraph*{Eliminating $u^a$ \label{sec:eliminating_u}}

Insertion of (\ref{uprel}) into the equation of motion for the spin (\ref{pap2}) yields:
\begin{eqnarray}
\frac{\delta S^{ab}}{d s} = \frac{2 S^{c[a} p^{b]}}{p^f f_f} \frac{\delta f_c}{d s} - \frac{4}{3} W^{[a}_f W^{b]}_g R^f{}_{cde} J^{gcde}.
\end{eqnarray}
Here we defined the projector $W^a_b$ with respect to the vector entering the supplementary condition (\ref{general_supp}), i.e.
\begin{eqnarray}
W^a_b := \delta^a_b - \frac{p^a f_b}{p^f f_f},	\quad f_a W^a_b = 0, \\
\quad W^a_b p^b = 0, \quad W^a_c S^{cb} = S^{ab}. \label{projector_def}
\end{eqnarray}
Now the spin dynamics is fixed, provided that $\frac{\delta f_a}{d s}$ is meaningful and $J^{abcd}$ is somehow given. This spin equation of motion also manifestly preserves the spin supplementary condition.

When $u^a$ is replaced in favor of $p^a$, the following relation between the velocity and the momentum is often more useful than (\ref{uprel}):
\begin{eqnarray}
u^a \stackrel{\ast}{=} \hat{p}^a + \frac{2 S^{ac} S^{de} R_{decb}}{4 \underline{m}^2 + S^{cd} S^{ef} R_{cdef}} \hat{p}^b \label{mom_vel_relation},
\end{eqnarray}
where 
\begin{eqnarray}
\hat{p}^a &:=& \frac{m}{\underline{m}^2} p^a - \frac{1}{\underline{m}^2} \frac{\delta}{ds} \left(S^{ab} p_b \right) - \frac{4}{3 \underline{m}^2} R^{[a}{}_{cde} J^{b]cde} p_b \nonumber \\
 &&+ \frac{1}{6 \underline{m}^2} S^{ab} \nabla_b R_{cdef} J^{cdef}, \label{def_hat_mom} \\
 &=& \frac{m}{\underline{m}^2} p^a - \frac{S^{ab}}{\underline{m}^2} \frac{\delta p_b}{d s}  + \frac{1}{6 \underline{m}^2} S^{ab} \nabla_b R_{cdef} J^{cdef} \nonumber \\ 
 && - \frac{4}{3 \, p^g f_g} \hat{\rho}_f^a R^{[f}{}_{cde} J^{b]cde} f_b +\frac{\hat{\rho}_b^a S^{bc}}{p^f f_f} \frac{\delta f_c}{d s}. 
\end{eqnarray}
In the last equation we introduced $\hat{\rho}_a^b := \delta_a^b - p_a p^b / \underline{m}^2$. The derivation of relation (\ref{mom_vel_relation}) at the quadrupole order is analogous to the one at the pole-dipole order given in \cite{Obukhov:Puetzfeld:2011:1}. Note that (\ref{mom_vel_relation}) is only valid, if the supplementary condition (\ref{general_supp}) holds.

The mass quantity $m := p_a u^a$ must be obtained from (\ref{mom_vel_relation}) and $u^a u_a = 1$, as $s$ is the proper time in our case. But one may as well use a different convention for the parameter $s$, namely $u_a U^a = 1$ where $U^a := p^a / \underline{m}$. Then it holds $m = \underline{m}$, which makes the relation between $u^a$ and $p^a$ fully explicit, see also \cite{Ehlers:Rudolph:1977}. Now one can insert (\ref{mom_vel_relation}) into (\ref{pap1}), which finally eliminates $u^a$ from the equations of motion. If desired, one can further decompose (\ref{pap1}) into an equation for $U^a$ and $\underline{m}$, but we will not explicitly follow this approach here.

\paragraph*{Eliminating $p^a$ \label{sec:eliminating_p}}

Equation (\ref{uprel}) can be written as:
\begin{equation}
p^a = \frac{1}{u^f f_f} \left[ p^b f_b u^a - S^{ab} \frac{\delta f_b}{d s} + \frac{4}{3} f_b R^{[a}{}_{cde} J^{b]cde} \right].
\end{equation}
Insertion into the equation of motion for the spin (\ref{pap2}) yields:
\begin{eqnarray}
\frac{\delta S^{ab}}{d s} = \frac{2 S^{c[a} u^{b]}}{u^f f_f} \frac{\delta f_c}{d s} - \frac{4}{3} X^{[a}_f X^{b]}_g R^f{}_{cde} J^{gcde},
\end{eqnarray}
where the projector $X^a_b$ is now given by
\begin{eqnarray}
X^a_b := \delta^a_b - \frac{u^a f_b}{u^f f_f},	\quad f_a X^a_b = 0, \\
\quad X^a_b u^b = 0, \quad X^a_c S^{cb} = S^{ab}. \label{projector_X_def}
\end{eqnarray}
As in the previous case this spin equation of motion manifestly preserves the spin supplementary condition.

With the help of the relations\footnote{Here $\rho^a_b:=\delta^a_b - u^a u_b$ denotes the projector with respect to the velocity as usual.}
\begin{eqnarray}
f_a p^a &= m f_a u^a + S^{ab} u_a \frac{\delta f_b}{d s} + \frac{4}{3} f_a u_{b} R^{[a}{}_{cde} J^{b]cde} ,\\
p^a &= m u^a	- \frac{\rho^a_b S^{bc}}{u^f f_f} \frac{\delta f_c}{d s} + \frac{4}{3} \rho^a_b \frac{f_g}{u^f f_f} R^{[b}{}_{cde} J^{g]cde}, \label{pu}
\end{eqnarray}
we can orthogonally decompose (\ref{pap1}) with respect to $u^{a}$ into an equation for $u^{a}$, and into an equation for $m$:
\begin{eqnarray}
m \frac{\delta u_a }{d s} &=& \frac{1}{2} R_{abcd} u^{b} S^{cd}	+ \rho_{ag} \frac{\delta}{d s} \left[ \frac{\rho^g_b S^{bc}}{u^f f_f} \frac{\delta f_c}{d s}  \right. \nonumber \\
&& \left. - \frac{4}{3} \rho^g_b \frac{f_h}{u^f f_f} R^{[b}{}_{cde} J^{h]cde} \right]	+ \frac{1}{6} \rho_a^f \nabla_f R_{bcde} J^{bcde}, \nonumber \\ 
&& \label{decomp_1}\\
\frac{d m}{d s} &=& 	- \frac{\delta u_b}{d s} \frac{S^{bc}}{u^f f_f} \frac{\delta f_c}{d s}	+ \frac{\delta u_b}{d s} \frac{4}{3} \frac{f_g}{u^f f_f} R^{[b}{}_{cde} J^{g]cde}	\nonumber \\
&& + \frac{1}{6} \frac{\delta R_{bcde}}{d s} J^{bcde}. \label{decomp_2}
\end{eqnarray}
Obviously $u_a u^a=1$ is preserved. Observe that we have eliminated $p^a$ from all equations of motion.

\section{Spinning test bodies in a Kerr background}\label{orbits_sec}

In the following, we are going to study test bodies endowed with spin in the field of a rotating source described by the Kerr metric. 

\subsection{Kerr metric}

In Boyer-Lindquist coordinates $(t,r,\theta,\phi)$, the Kerr metric takes the form:
\begin{eqnarray}
ds^2&=&\left(1-\frac{2Mr}{\rho^2}\right)dt^2+\frac{4aMr {\rm sin}^2\theta}{\rho^2} dtd\phi - \frac{\rho^2}{\Delta} dr^2 \nonumber \\ 
&&- \rho^2d\theta^2 -{\rm sin}^2\theta \left(r^2+a^2+\frac{2a^2Mr {\rm sin}^2\theta}{\rho^2} \right)d\phi^2, \nonumber \\ \label{kerr_metric}
\end{eqnarray}
where $M$ is the mass, $a$ the Kerr parameter, and 
\begin{eqnarray}
\Delta = r^2 - 2 M r + a^2, \quad \rho^2 = r^2 + a^2 {\rm cos}^2\theta.
\end{eqnarray}
The Kerr metric allows for two Killing vector fields given by:
\begin{eqnarray}
{\underset{_E}\xi^a}=\delta^a_t, \quad \quad {\underset{_J}\xi^a}=\delta^a_\phi. \label{kerr_killing}
\end{eqnarray}
Furthermore, we have
\begin{eqnarray}
\sqrt{-g}:=\sqrt{-{\rm det}\left(g_{ab}\right)}=\rho^2 {\rm sin}\theta. \end{eqnarray}

\subsection{Equatorial orbits for aligned spin}\label{sub_eq_orbits}

In the following we are going to focus on equatorial orbits, i.e.\
\begin{eqnarray}
\theta = \frac{\pi}{2}, \quad \quad p^{\theta}=0, \label{eq_orbits_def}
\end{eqnarray} 
and the case of aligned spin, with respect to the rotating background source, defined by
\begin{eqnarray}
S^{a\theta} = 0. \label{aligneds}
\end{eqnarray}
The self-consistency of this configuration was shown in \cite{Rasband:1973, Hojman:Hojman:1977}, see also Appendix \ref{sec_existence} of the present paper.
These assumptions leave us with the six remaining components $\{p^t,p^r,p^\phi,S^{tr},S^{r\phi},S^{\phi t}\}$, which can be determined from the following set of equations
\begin{eqnarray}
S^{ab}p_b &=& 0, \quad S_{ab} S^{ab} = 2 S^2, \quad \quad p_a p^a = \underline{m}^2, \nonumber \\
E &=& p_a {\underset{_E}\xi^a} + \frac{1}{2} S^{a}{}_b \nabla_a {\underset{_E}\xi^b}, \nonumber \\
-J &=& p_a {\underset{_J}\xi^a} + \frac{1}{2} S^{a}{}_b \nabla_a {\underset{_J}\xi^b}, \label{six_equations}
\end{eqnarray}
in terms of the mass $\underline{m}$, the spin length $S$, the energy $E$, and the angular momentum $J$. Remember the corresponding Komar quantities of the Kerr background at this point. The one belonging to the Killing vector $\underset{_E}\xi^a$ is just the black hole mass (or energy), the other one, belonging to ${\underset{_J}\xi^a}$, yields the angular momentum. This motivates to call $E$ and $J$ the energy and angular momentum of the test body, as they are based on the same isometries, or Killing vectors. Note that from here on we use the Tulczyjew condition, i.e.\ the first equation in (\ref{six_equations}), as supplementary condition in our multipole formalism. For this condition, we define the spin vector $S^a$ as follows:  
\begin{eqnarray}
S^a&=&\frac{1}{2 \underline{m}} \eta^{abcd} p_b S_{cd} = \frac{1}{2 \underline{m} \sqrt{-g}} \varepsilon^{abcd} p_b S_{cd}, \nonumber \\
S^{ab}&=&\frac{1}{\underline{m}} \eta^{abcd} p_c S_{d} = \frac{1}{\underline{m} \sqrt{-g}} \varepsilon^{abcd} p_c S_{d}. \label{S_vector}
\end{eqnarray}
From this definition and the assumptions in (\ref{eq_orbits_def}) and (\ref{aligneds}) it becomes clear, that $S^a$ has only one non-vanishing component, i.e.\
\begin{eqnarray}
S^a = S^\theta \delta^a_\theta.
\end{eqnarray}
Using the relation $2 S_a S^a = - S_{ab} S^{ab}$ together with the definition of the spin length from (\ref{six_equations}), we obtain 
\begin{equation}
- S^{\theta} = \pm S / \sqrt{- g_{\theta\theta}} .
\end{equation}
Notice that for usual spherical coordinates $\partial_{\theta}$ points in the opposite direction as $\partial_z$ in the equatorial plane. Therefore $S^{\theta} < 0$ corresponds to a spin aligned to $\partial_z$. Further discussion of the spin orientation is given in Appendix \ref{sec_orientation}. As usual we absorb the sign by allowing for negative spin length $S$ in the following. This in turn allows us to express the components of the spin tensor in terms of the spin length, i.e.,
\begin{eqnarray}
S^{rt}&=&-\frac{S \sqrt{-g_{\theta \theta}}}{\underline{m} \sqrt{-g}} p_\phi=-\frac{S p_\phi}{\underline{m} r}, \nonumber \\
S^{\phi t}&=&\frac{S \sqrt{-g_{\theta \theta}}}{\underline{m} \sqrt{-g}} p_r=\frac{S p_r}{\underline{m} r}, \nonumber \\
S^{\phi r}&=&-\frac{S \sqrt{-g_{\theta \theta}}}{\underline{m} \sqrt{-g}} p_t=-\frac{S p_t}{\underline{m} r}, 
\end{eqnarray}
where in the last step we took into account that we are in the equatorial plane.

The definitions of $E$ and $J$ in (\ref{six_equations}) allow us to express $p_t$ and $p_\phi$ in terms of the constants of motion, i.e.\
\begin{eqnarray}
p_t&=&\frac{E-\frac{MS}{\underline{m}r^3}\left(J-aE\right)}{1-\frac{MS^2}{\underline{m}^2r^3}}, \label{p_t_explicit} \\
p_\phi&=&\frac{-J-\frac{aMS}{\underline{m}r^3}\left[aE\left(1-\frac{r^3}{a^2M}\right)-J\right]}{1-\frac{MS^2}{\underline{m}^2r^3}} \label{p_phi_explicit}.
\end{eqnarray}
With the last two expressions and the definition of the mass $\underline{m}$ from (\ref{six_equations}) we are able to express the radial component of the linear momentum, $p^r$, in terms of the constants of motion, the test body mass, the test body spin, and the parameters characterizing the spacetime, i.e.\
\begin{eqnarray}
(p^r)^2 &=& A_0 \left(A_1 J^2 + A_2 JE + A_3 E^2 + A_4 \right). \label{p_r_explicit}
\end{eqnarray}
Here the functions $A_0=A_0\left(M,\underline{m},S,r\right)$, $A_{1,\dots,4}=A_{1,\dots,4}\left(M,\underline{m},S,r,a\right)$ are explicitly given by:
\begin{eqnarray}
A_0&=&\frac{\underline{m}^2M}{\left(S^2M-\underline{m}^2r^3\right)^2}, \\
A_1&=&S^2M+2\underline{m}rSa+2r^3\underline{m}^2-\frac{\underline{m}^2r^4}{M}, \\
A_2&=&\frac{2\underline{m}r^4S}{M}-6r^3S\underline{m}-4a\underline{m}^2r^3-2S^2aM \nonumber \\
&&-2S^2ar-4\underline{m}rSa^2, \\
A_3&=&2a^2\underline{m}^2r^3+2a^3rS\underline{m}+2S^2r^3+6r^3Sa\underline{m}+\frac{r^6\underline{m}^2}{M}\nonumber  \\
&&+2S^2ra^2+S^2Ma^2-\frac{S^2r^4}{M}+\frac{a^2r^4\underline{m}^2}{M}, \\
A_4&=&2r^3S^2\underline{m}^2-S^4M-\frac{r^6\underline{m}^4}{M}+\frac{2M^2S^4}{r}-4Mr^2S^2\underline{m}^2 \nonumber\\
&&+2r^5\underline{m}^4-\frac{a^2S^4M}{r^2}+2a^2S^2\underline{m}^2r-\frac{a^2\underline{m}^4r^4}{M}.
\end{eqnarray}
An alternative form of $p^r$ is the following one:
\begin{equation}
\left(\frac{p^r}{\underline{m}}\right)^2 = \frac{1}{\sigma} \left[ \alpha \frac{E^2}{\underline{m}^2} - 2 \beta \frac{J E}{\underline{m}^2 r} + \gamma \frac{J^2}{\underline{m}^2 r^2} - \delta \right] , \label{pot1st}
\end{equation}
\begin{eqnarray}
\alpha &=& \left[ 1 + \frac{a^2}{r^2} + \left(1+ \frac{M}{r}\right) \frac{a S}{\underline{m} r^2} \right]^2 \nonumber \\
&&- \frac{\Delta}{r^2} \left(\frac{a}{r} + \frac{S}{\underline{m} r}\right)^2 , \\
\beta &=& \left[ 1 + \frac{a^2}{r^2} + \left(1+ \frac{M}{r}\right) \frac{a S}{\underline{m} r^2} \right]\left(\frac{a}{r} + \frac{M S}{\underline{m} r^2}\right) \nonumber \\
	&&- \frac{\Delta}{r^2} \left(\frac{a}{r} + \frac{S}{\underline{m} r}\right) , \\
\gamma &=& \left(\frac{a}{r} + \frac{M S}{\underline{m} r^2}\right)^2 - \frac{\Delta}{r^2} , \\
\delta &=& \frac{\Delta}{r^2} \sigma , \quad
\sigma = \left(1 - \frac{M S^2}{\underline{m}^2 r^3}\right)^2. \label{potlast}
\end{eqnarray}

\subsection{Effective potential (pole-dipole case)}\label{effective_potential_sec}

\begin{figure}
\begin{center}
\subfigure[$\quad$ Parameter choice $a=0.5 M$, $S=\underline{m}M$, $J=4\underline{m}M$.]
{\includegraphics{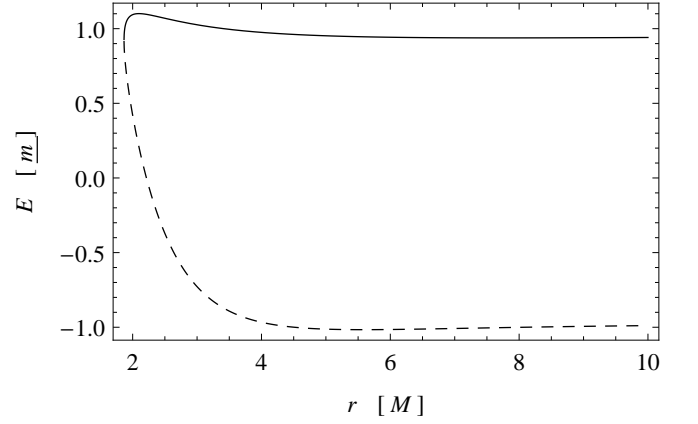}\label{fig_potential_spinning_1}}
\subfigure[$\quad$ Parameter choice $a=0.5 M$, $S=-\underline{m}M$, $J=4\underline{m}M$.]
{\includegraphics{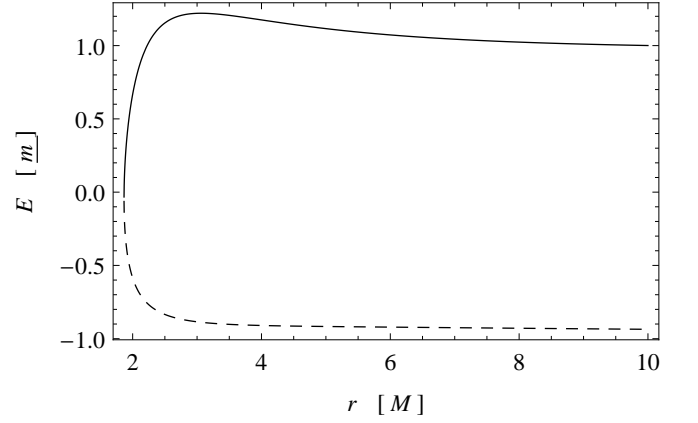}\label{fig_potential_spinning_2}}
\caption{Energy $E=U_\pm$, in units of $\underline{m}$, as a function of the radius $r$, in units of $M$.\label{fig_potential_spinning_both}}
\end{center}
\end{figure}

We may define the functions $U_{+}$ and $U_{-}$ by
\begin{equation}
\left(p^r\right)^2 = \frac{\alpha}{\sigma} (E - U_{+}) (E - U_{-}) ,
\end{equation}
so that
\begin{equation}
\frac{U_{\pm}}{\underline{m}} = \pm \sqrt{\frac{\delta}{\alpha} + \frac{M^2}{r^2} \frac{\beta^2 - \alpha \gamma}{\alpha^2} \frac{J^2}{\underline{m}^2 M^2}}	+ \frac{M}{r} \frac{\beta}{\alpha} \frac{J}{\underline{m} M}. \label{U_definition}
\end{equation}
For $p^r$ to be a real number we need to have both $E \leq U_{+}$ and $E \leq U_{-}$, or both $E \geq U_{+}$ and $E \geq U_{-}$ (under the assumption\footnote{We numerically checked that $\alpha > 0$ in the regime $M < r < \infty$, $-M \leq a \leq M$, $- M \underline{m} \leq S \leq M \underline{m}$.} that $\alpha > 0$). For usual cases (and positive energy $E \geq 0$) the important relation is $E \geq U_{+}$. This justifies to call $U_{+}$ effective potential: The object can only move in the region where $E \geq U_{+}$ and the turning points are given by $E = U_{+}$, because then $p^r = 0$ -- which implies $u^r = 0$, see Appendix \ref{sec_existence}. Therefore the minimum of $U_{+}$,
\begin{equation}
\frac{\partial U_{+}}{\partial r} = 0 , \label{Ucirc}
\end{equation}
together with $E = U_{+}$ defines circular orbits. Thus, one can easily find circular orbit solutions by solving (\ref{Ucirc}) for the radial coordinate $r$ in terms of constants of motion and the other constant parameters. For this class of solutions the energy is constrained by $E = U_{+}$.

We have plotted the solutions for the energy $E = U_{\pm}$ for different spin configurations in fig.\ \ref{fig_potential_spinning_both}, these can be directly compared to the ones found in \cite{Suzuki:Maeda:1998}. 

\section{Explicit quadrupole model}\label{quadrupole_model_sec}

We are interested in the models for quadrupole deformations induced by spin given by \cite[eqs.\ (1) and (16)]{Porto:Rothstein:2008:2} and for \emph{adiabatic} tidal quadrupole deformations given by \cite[(19)]{Goldberger:Rothstein:2006:2}, or by \cite[(5)]{Damour:Nagar:2009}. These models are an extension of the point-mass action, see Appendix \ref{sec_quadrupole_action}, and the corresponding equations of motion have been worked out in a general fashion already in \cite{Bailey:Israel:1975}. In particular, equation \cite[(19)]{Bailey:Israel:1975} provides a formula for the quadrupole $J^{abcd}$. Other recent treatments (not based on an action principle) of tidal effects in the context of General Relativity can be found in \cite{Ferrari:Gualtieri:Pannarale:2009, Binnington:Poisson:2009, Ferrari:Gualtieri:Maselli:2012}. Inspired by the abovementioned models we now consider
\begin{equation}
J^{abcd} = - \frac{m}{\underline{m}} \left[ \frac{1}{\underline{m}} p^{[a} Q^{b]cd} + \frac{1}{\underline{m}} p^{[d} Q^{c]ba}	+ \frac{3}{\underline{m}^2} p^{[a} Q^{b][c} p^{d]} \right], \label{Jex1st}
\end{equation}
where
\begin{eqnarray}
Q^{ab} &=& c_{ES^2} S^a{}_{e} S^{be} - \mu_2 E^{ab} , \nonumber \\
Q^{bcd} &=& - \frac{2 \sigma_2}{\underline{m}} \eta^{dc}{}_{ea} p^e B^{ba} , \nonumber \\
E_{ab} &=& \frac{1}{\underline{m}^2} R_{acbd} p^c p^d,  \nonumber \\
B_{ab} &=& \frac{1}{2 \underline{m}^2} \eta_{aecd} R_{bf}{}^{cd} p^e p^f . \nonumber
\end{eqnarray}
The quantities $c_{ES^2}$, $\mu_2$, and $\sigma_2$ are assumed to be constants, and parameterize quadrupole deformations induced by the spin and by tidal forces of the spacetime. Furthermore,  $E^{ab}$ represents the gravito-electric tidal field, and $B^{ab}$ the gravito-magnetic (frame-dragging) tidal field, see, e.g., \cite{Nichols:etal:2011}. (The convention for $B^{ab}$ used in \cite{Damour:Nagar:2009,Bini:Damour:Faye:2012} differs from the one adopted in the present paper by a factor of two.) Notice that the overall factor of $m / \underline{m}$ in $J^{abcd}$ makes the equations of motion reparametrization invariant. We choose the spin supplementary condition of Tulczyjew,
\begin{equation}
S^{ab} p_b = 0 ,
\end{equation}
as this condition is most convenient for the derivation of an effective potential (similar to the pole-dipole case).

\subsection{Spin length}

For the explicit model given by (\ref{Jex1st}) we have
\begin{eqnarray}
S \frac{d S}{d s} &=& - \frac{1}{6 \sigma_2} S_{ab} ( Q^a{}_{de} Q^{bde} + 2 Q_{dc}{}^a Q^{dcb} ) \nonumber \\
&&	+ \mu_2 S_{ab} E^a{}_d E^{bd}	+ c_{ES^2} E^{ad} S_{ab} S^{be} S_{ed} ,
\end{eqnarray}
where (\ref{dSds}) and the Tulczyjew condition were used. Furthermore, one has a symmetry of the combinations
\begin{equation}
E^{ab} , \quad E^a{}_d E^{bd} , \quad Q^a{}_{de} Q^{bde} , \quad Q_{dc}{}^a Q^{dcb} ,
\end{equation}
and the antisymmetry of
\begin{equation}
S_{ab} , \quad S_{ad} S^{de} S_{eb} ,
\end{equation}
under exchange of $a$ and $b$. It immediately follows that the spin length is conserved,
\begin{equation}
\frac{d S}{d s} = 0 .
\end{equation}
This conservation law could be expected, as the action given in Appendix \ref{sec_quadrupole_action} -- which served as an inspiration for the present quadrupole model -- possesses a symmetry under rotations of the body-fixed frame. But the action is only consistent with the spin supplementary condition to a certain power in spin, whereas the conservation found here made no reference to such an approximation.

\subsection{Mass-like quantity} \label{sec:quadmass}

How a possibly conserved mass-like quantity is related to the usual masses $\underline{m}$, or $m$, crucially depends on an explicit model for the quadrupole. In the present paper a conserved mass-like quantity is related to a parameter within the underlying action, which is a constant by \emph{assumption}. However, the action is only consistent with the spin supplementary condition in an approximate sense, which translates here to the fact that we were only able to find a mass-like quantity which is approximately constant.

We therefore introduce a multipole counting scheme by a symbol $\epsilon$ of the form
\begin{eqnarray}
m &=& \Order{(\epsilon^0)} , \quad u^a = \Order{(\epsilon^0)} , \quad \frac{\delta p^a}{d s} = \Order{(\epsilon^1)} , \\
S^{ab} &=& \Order{(\epsilon^1)}, \quad \frac{\delta S^{ab}}{d s} = \Order{(\epsilon^2)} , \quad J^{abcd} = \Order{(\epsilon^2)} .
\end{eqnarray}
We will neglect $\Order{(\epsilon^3)}$ terms in the following. Under these conditions, (\ref{dmbdt}) simplifies to
\begin{equation}
\frac{d \underline{m}}{d s} = \frac{1}{6} \frac{\delta R_{abcd}}{d s} J^{abcd} + \Order{(\epsilon^3)} .
\end{equation}
We define the mass-like parameter $\mu$ given by
\begin{equation}
\mu := \underline{m}	+ \frac{c_{ES^2}}{2} E_{ab} S^a{}_{c} S^{cb} + \frac{\mu_2}{4} E_{ab} E^{ab} + \frac{2\sigma_2}{3} B_{ab} B^{ab} . \label{const_mass}
\end{equation}
This mass is indeed approximately conserved,
\begin{equation}
\frac{d \mu}{d s} = 0 + \Order{(\epsilon^3)}.
\end{equation}
Notice that $p_a = m u_a + \Order{(\epsilon^2)}$ and (\ref{mass_relation}) lead to $\underline{m}^2 = m^2 + \Order(\epsilon^3)$. This implies that the Tulczyjew and the Frenkel condition are actually equivalent within this approximation.

One can guess (\ref{const_mass}) by realizing that -- because of reparametrization invariance -- the Lagrangian (or Routhian $R_M$) must be equal to
\begin{equation}
R_M = u^a p_a \equiv m = \underline{m} + \Order(\epsilon^3) .
\end{equation}
In fact, (\ref{const_mass}) is identical to (\ref{Routhian}) with $S^{ab} u_b \approx S^{ab} p_b = 0$ and $u^2 = 1$ inserted. 

\subsection{Dimensionless parameters and realistic values}

Before we work out the effective potential for the quadrupole case, it is useful to introduce dimensionless variables\footnote{Note that in our units $c=1$, so $[G] = \text{m}/\text{kg}$, see also section \ref{dimension_acronyms_app} for an overview.},
\begin{eqnarray}
\hat{r} := \frac{r}{M}, \quad \hat{J} := \frac{J}{M \underline{m}} , \quad \hat{E} := \frac{E}{\mu} , \\
\hat{a}_1 := \frac{a}{M}, \quad \hat{a}_2 := \frac{S}{G \mu^2}.
\end{eqnarray}
Astrophysically reasonable values for the dimensionless spin variables are given by
\begin{equation}
| \hat{a}_1 | \lesssim 1, \qquad | \hat{a}_2 | \lesssim 1.
\end{equation}
Notice that
\begin{equation}
\frac{S}{M \underline{m}} = q \hat{a}_2 + \Order(\epsilon^3), \quad \text{with } q := \frac{G \mu}{M},
\end{equation}
so spin effects are strongly suppressed for extreme mass ratios $q \ll 1$.

The tidal deformation parameters $\mu_2$ and $\sigma_2$ are usually made dimensionless with the help of the area radius $R$ of the object, see \cite[(48) and (72)]{Damour:Nagar:2009:4}, i.e.,
\begin{equation}
k_2 := \frac{3 G \mu_2}{2 R^5} , \qquad j_2 := \frac{48 G \sigma_2}{R^5} .
\end{equation}
These parameters can be obtained by matching predictions derived from the effective action (Appendix \ref{sec_quadrupole_action}) to solutions of the field equations describing a \emph{single} object (in asymptotically flat spacetime). This illustrates the phenomenological character of the quadrupole model. Realistic values for neutron stars are $k_2 \sim 0.1$ and $j_2 \sim -0.02$, while for black holes $\mu_2 \sim 0$, and $\sigma_2 \sim 0$, see \cite{Damour:Nagar:2009:4,Binnington:Poisson:2009}. For tidal deformations of black holes see also \cite{Kol:Smolkin:2012,Comeau:Poisson:2009,Vega:Poisson:Massey:2011} and references therein. An estimate for white dwarfs is $k_2 \sim 0.01$ \cite{Prodan:Murray:2012}. For convenience we also define a dimensionless radius $\hat{R}$,
\begin{equation}
\hat{R} := \frac{R}{G \mu} ,
\end{equation}
which is just the inverse of the compactness of the test body. Finally, we introduce
\begin{equation}
C_{ES^2} := \mu c_{ES^2} ,
\end{equation}
which for neutron stars is of the order $C_{ES^2} \sim 5$ \cite{Laarakkers:Poisson:1999} and for black holes it holds $C_{ES^2} = 1$. Now we are ready to express the effective potential in terms of dimensionless quantities only.

\subsection{Effective potential (quadrupole case)}

Let us recall how the effective potential in the pole-dipole case was derived. Under the assumption of equatorial orbits (\ref{eq_orbits_def}) and aligned spin (\ref{aligneds}), the six independent equations (\ref{six_equations}) are solved for the six variables \{$p^t$, $p^r$, $p^\phi$, $S^{tr}$, $S^{r\phi}$, $S^{\phi t}$\}. All components of $p^a$ and $S^{ab}$ can then be expressed in terms of \{$r$, $E$, $J$, $S$, $a$, $M$, $\underline{m}$\}. Besides the rather technical issue of proofing the existence of equatorial orbits for our quadrupole model (see Appendix \ref{sec_existence}), there are no changes in this part of the derivation. The spin supplementary condition in (\ref{six_equations}) is still valid/chosen, the second and third equation in (\ref{six_equations}) are just the definitions of spin length $S$ and dynamical mass $\underline{m}$, and it was shown in \cite[p.\ 210]{Ehlers:Rudolph:1977} that energy $E$ and total angular momentum $J$ of the test body are given by the same expressions as in the pole-dipole case, even if generic multipole corrections are included. The bottom line is that the solutions for $p^a$ and $S^{ab}$ found in sec.\ \ref{sub_eq_orbits} are still valid. The effective potential was defined as value of $E$ for which $p^r$ vanishes. Notice that $p^r = 0$ implies that the orbit either has a turning point or is circular, see Appendix \ref{sec_existence}.

The most important application of the effective potential is to find circular orbit solutions. These are given by a minimum of the effective potential in the radial coordinate $r$. We therefore need to work out the $r$-dependence of the quantities entering the effective potential when our quadrupole model is used. The spin length $S$ is still constant, so the only correction is coming from the dynamical mass $\underline{m}$. Indeed, $\underline{m}$ is not a constant any more and given by (\ref{const_mass}). More generally, we conclude that whenever the supplementary condition $S^{ab} p_{b} = 0$ is used, and the spin length is constant, the quadrupole and higher multipoles enter the effective potential via a $r$-dependence of the dynamical mass $\underline{m}$ only. This is similar to canonical theories for self-gravitating bodies, in which just the mass-shell constraint is modified, see \cite[eq.\ (5.28)]{Steinhoff:2011}.

Inserting the solutions for $p^a$ and $S^{ab}$ from sec.\ \ref{sub_eq_orbits} into (\ref{const_mass}) and writing the result in terms of dimensionless quantities, we obtain
\begin{eqnarray}
\frac{\underline{m}}{\mu} &=& 1 - \frac{C_{ES^2} q^2 \hat{a}_2^2}{2 \hat{r}^3} \left[ 1 + 3 \left( \frac{\hat{J} - \hat{a}_1 \hat{E}}{\hat{r}} \right)^2 \right] \nonumber \\
&&- \frac{k_2 q^4 \hat{R}^5}{\hat{r}^6} \left[ 1 + 3 \left( \frac{\hat{J} - \hat{a}_1 \hat{E}}{\hat{r}} \right)^2 + 3 \left( \frac{\hat{J} - \hat{a}_1 \hat{E}}{\hat{r}} \right)^4 \right] \nonumber \\
&&- \frac{j_2 q^4 \hat{R}^5}{4 \hat{r}^6} \left[ \left( \frac{\hat{J} - \hat{a}_1 \hat{E}}{\hat{r}} \right)^2 + \left( \frac{\hat{J} - \hat{a}_1 \hat{E}}{\hat{r}} \right)^4 \right]\nonumber \\
&&+ \Order(\epsilon^3). \label{mquad}
\end{eqnarray}
Consistent with our approximation we have replaced $\underline{m}$ by $\mu$ on the right hand side and neglected higher orders of $S$. Similarly one may replace $\hat{E}$ by the $\epsilon^0$-order solution for the effective potential. Otherwise the equation $p^r = 0$ with
\begin{equation}
\left(\frac{p^r}{\mu}\right)^2 = \frac{1}{\sigma}	\left[ \alpha \hat{E}^2 - 2 \beta \frac{\hat{J}}{\hat{r}} \hat{E} + \gamma \frac{\hat{J}^2}{\hat{r}^2} - \delta \frac{\underline{m}^2}{\mu^2} \right] ,
\end{equation}
would contain higher orders of $\hat{E}$, and it would be difficult to solve for the effective potential analytically. If one treats $\hat{E}$ in (\ref{mquad}) perturbatively, then one still has only two solutions,
\begin{equation}
\hat{U}_{\pm} = \pm \sqrt{ \frac{\delta}{\alpha} \frac{\underline{m}^2}{\mu^2} + \frac{\beta^2 - \alpha \gamma}{\alpha^2} \frac{\hat{J}^2}{\hat{r}^2} }	+ \frac{\beta}{\alpha} \frac{\hat{J}}{\hat{r}}, \label{pot_quad_rewritten}
\end{equation}
to $p^r = 0$. Here $\hat{U}_{\pm} = U_{\pm} / \mu$. For completeness, we also give the auxiliary variables in (\ref{pot_quad_rewritten}) in terms of dimensionless quantities,
\begin{eqnarray}
\alpha &=& \left[ 1 + \frac{\hat{a}_1^2}{\hat{r}^2} + \left(1+ \frac{1}{\hat{r}}\right) \frac{q \hat{a}_1 \hat{a}_2}{\hat{r}^2} \right]^2  \nonumber \\ 
&& - \frac{\hat{\Delta}}{\hat{r}^2} \left(\frac{\hat{a}_1}{\hat{r}} + \frac{q \hat{a}_2}{\hat{r}}\right)^2 + \Order(\epsilon^3) , \\
\beta &=& \left[ 1 + \frac{\hat{a}_1^2}{\hat{r}^2} + \left(1 + \frac{1}{\hat{r}}\right) \frac{q \hat{a}_1 \hat{a}_2}{\hat{r}^2} \right] \left(\frac{\hat{a}_1}{\hat{r}} + \frac{q \hat{a}_2}{\hat{r}^2}\right) \nonumber \\ 
	&& - \frac{\hat{\Delta}}{\hat{r}^2} \left(\frac{\hat{a}_1}{\hat{r}} + \frac{q \hat{a}_2}{\hat{r}}\right)	+ \Order(\epsilon^3) , \\
\gamma &=& \left(\frac{\hat{a}_1}{\hat{r}} + \frac{q \hat{a}_2}{\hat{r}^2}\right)^2 - \frac{\hat{\Delta}}{\hat{r}^2} + \Order(\epsilon^3) , \\
\delta &=& \frac{\hat{\Delta}}{\hat{r}^2} \sigma , \\
\sigma &=& \left(1 - \frac{q^2 \hat{a}_2^2}{\hat{r}^3}\right)^2 + \Order(\epsilon^3) , \\
\hat{\Delta} &=& \hat{r}^2 - 2 \hat{r} + \hat{a}^2 .
\end{eqnarray}
Notice that $\underline{m}$ appears here only within the $S$-dependent terms, where it can simply be substituted by the conserved mass-like parameter $\mu$ within the used approximation.

\subsection{Tidal Disruption}

When tidal forces become too large the test body can be disrupted. This limits the effects of tidal deformation on the effective potential. Equating the tidal force and the self-gravitational force of a non-rotating test body on its surface in Newtonian theory leads to
\begin{equation}
\frac{2 M}{G r^3}  R \lesssim \frac{\mu}{R^2} ,
\end{equation}
which provides a rough estimate for the orbital separation $r$ at which tidal disruption may become relevant. In dimensionless variables this reads
\begin{equation}
\hat{r}^3 \gtrsim 2 q^2 \hat{R}^3 .
\end{equation}

On the other hand, an estimate for (\ref{mquad}) by its leading post-Newtonian contribution, cf.\ sec.\ \ref{sec:PN}, reads (for the non-rotating case,
$\hat{a}_2 = 0$)
\begin{equation}
\frac{\underline{m}}{\mu} \approx 1 - \frac{k_2 q^4 \hat{R}^5}{\hat{r}^6} ,
\end{equation}
which limits the possible difference of the masses due to tidal disruption as
\begin{equation}\label{disrupt}
\frac{\mu - \underline{m}}{\mu} \lesssim \frac{k_2}{4} \frac{1}{\hat{R}} .
\end{equation}
This can be understood as follows. The dynamical mass $\underline{m}$ also includes the tidal interaction energy, which (for $k_2>0$) is negative, thus the dynamical mass is \emph{reduced}. This can be interpreted as a reduction of the gravitational potential energy (with respect to the ``external'' gravitational field) of the object due to its tidal deformation. If the tidal interaction energy reaches the order of $k_2 / (4 \hat{R})$, then the object is starting to get disrupted. This in turn limits the possible reduction of $\underline{m}$ due to tidal effects. Implications are discussed in the following.

Notice that the estimate in the present section is just Newtonian and thus might not be accurate for very compact objects like neutron stars; see, e.g., \cite{Ferrari:Gualtieri:Pannarale:2010} for this case.

\section{Binding energy}\label{sec_binding_energy}

In this section we explain how the gauge invariant relation between binding energy and total angular momentum can be obtained from the effective potential (for circular orbits and aligned spin). The various spin and quadrupole contributions to this relation are separated and compared against each other. 

The (conservative) self-force correction to the binding energy recently derived in \cite{LeTiec:Barausse:Buonanno:2012} is also included in our plots as a further reference point. Results from full numerical simulations can be found in \cite{Damour:Nagar:Pollney:Reisswig:2011}.

\subsection{Definition}

The effective potential $U_{+}$ takes the value of the constant energy $E$ of the test body in the case $p^r = 0$, e.g., for circular orbits (\ref{Ucirc}). We therefore define the binding energy as
\begin{equation}
e := \hat{U}_{+} - 1 , \label{def_e}
\end{equation}
under the condition that the orbit is circular (\ref{Ucirc}), or
\begin{equation}\label{circ_cond}
\frac{\partial e(\hat{r}, \hat{J})}{\partial \hat{r}} = 0 .
\end{equation}
This condition allows one to solve for the radial coordinate $\hat{r}$, which can subsequently be eliminated to arrive at the binding energy $e(\hat{J})$ as a function of the total angular momentum $\hat{J}$. This relation $e(\hat{J})$ is actually gauge invariant, which in the present context can be understood easily. That is, both the energy $E$ and the total angular momentum $J$ are scalars defined in a covariant manner based on the Killing vectors of the background Kerr spacetime, and circular orbits are a gauge independent concept.

Given the square root, and the high powers of $\hat{r}$ appearing in $U_{+}$, it is necessary to solve (\ref{circ_cond}) numerically. However, in order to better separate the different contributions to $e(\hat{J})$ and analyze their scaling behavior in $q$, $\hat{R}$, etc., we will expand (\ref{def_e}) and (\ref{circ_cond}) in the multipole counting parameter $\epsilon$ in the following. This allows one to solve (\ref{circ_cond}) for $\hat{r}$ order-by-order in $\epsilon$ \emph{analytically} if the background is Schwarzschild, i.e.\ $\hat{a}_1 = 0$. We will first consider the Schwarzschild case in the next section and come back to a Kerr spacetime in sec.\ \ref{sec_corr_in_Kerr}. But in general it is expected to be more accurate to not expand the square root appearing within the effective potential, as this better reflects the non-linear aspect of gravitational interaction. This is of particular importance if one extrapolates to comparable masses $q \sim 1$, but this is beyond the scope of the present work.

\subsection{Corrections in Schwarzschild spacetime\label{sec_corr_in_Schwarzschild}}

\subsubsection{Spin Effects}

There is a subtlety that needs to be discussed when looking at effects of the test body's spin $\hat{a}_2$. Namely, the total angular momentum also contains the spin of the test body. In canonical formulations, the total angular momentum is just the sum of the canonical orbital angular momentum $l_c$ and the spin of the object, i.e.,
\begin{equation}
l_c := \frac{1}{M \mu} (J-S) = \hat{J} - q \hat{a}_2. \label{deflc}
\end{equation}
The reason is that the total angular momentum generates rotations of the whole system on the phase space, which directly decompose into a rotation of the orbital variables (generated by $l_c$) and a rotation of the object (generated by the spin), see, e.g., \cite[sec.\ 4.1.2]{Steinhoff:2011}. If one now considers the Newtonian energy of point-masses
\begin{equation}
e_N(\hat{r}, l_c) = \frac{l_c^2}{2 \hat{r}^2} - \frac{1}{\hat{r}} , 
\end{equation}
and rewrites it in terms of the total angular momentum
\begin{equation}
e_N(\hat{r}, \hat{J}) = \frac{\hat{J}^2}{2 \hat{r}^2} - \frac{1}{\hat{r}} - \frac{\hat{a}_2 \hat{J} q}{\hat{r}^2} + \Order(\epsilon^2) ,
\end{equation}
an apparent spin-dependence of the binding energy arises. However, if one compares spinning and non-spinning systems at the same \emph{total} angular momentum $\hat{J}$, one is comparing them for different \emph{orbital} angular momentum, i.e., one is comparing them in different orbital configurations. The difference in their energies is just due to this change of the orbit, and not due to an actual spin interaction. There is nothing wrong here, this just shows that comparing $e(\hat{J})$ can be misleading. We will therefore use $e(l_c)$ for comparisons here, as for the same value of $l_c$ the system is approximately in the same orbital configuration. 

\begin{figure}
\begin{center}
\subfigure[$\quad$ Normalized curves for spin and self-force corrections.]
{\includegraphics{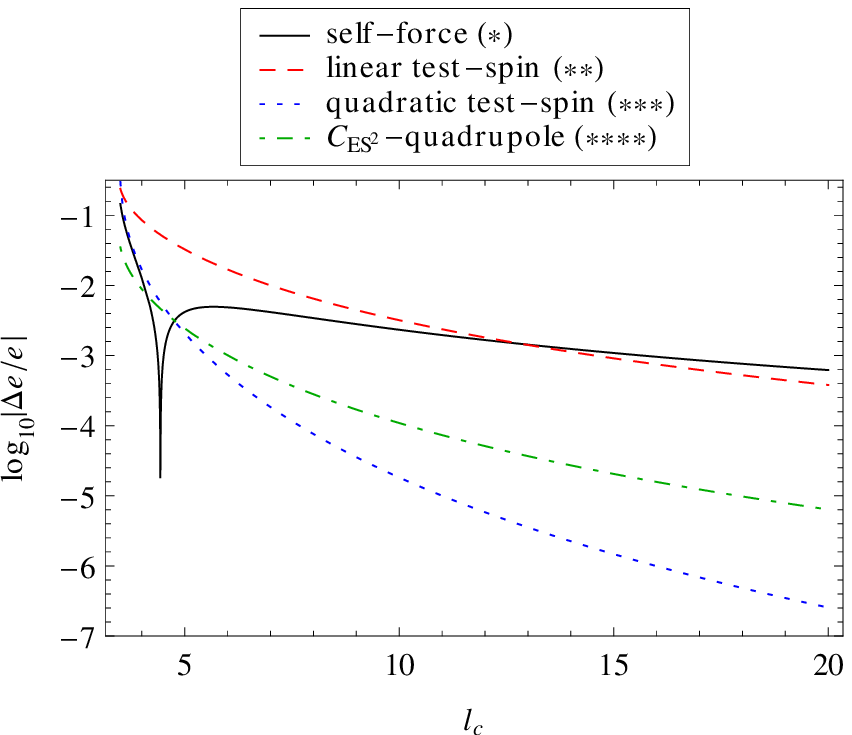}\label{spinPlot}}
\subfigure[$\quad$The same as \subref{spinPlot}, but zoomed in at small $l_c$.]
{\includegraphics{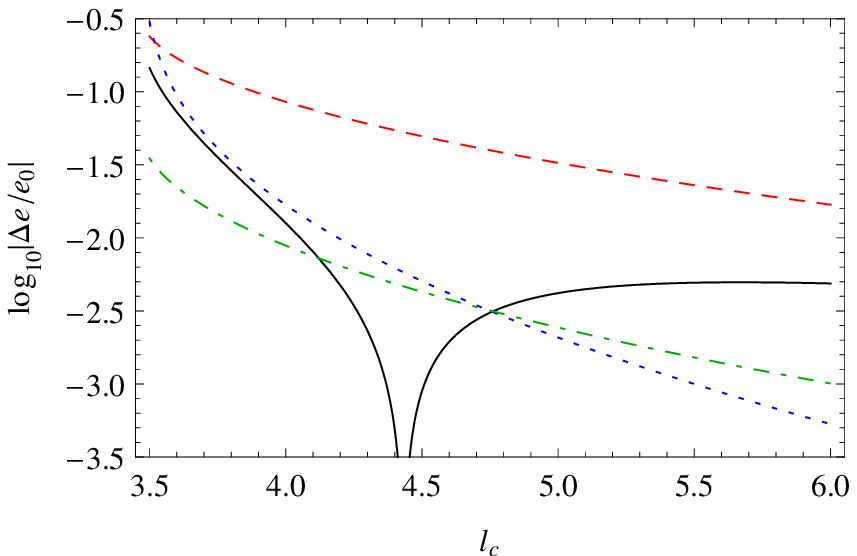}\label{spinPlotZoom}}
\subfigure[$\quad$The same as above, but the normalized curves are shifted down to a mass ratio $q=10^{-2}$ for $\hat{a}_2 = 1$ and $C_{ES^2} = 1$, which can occur for a black hole.]
{\includegraphics{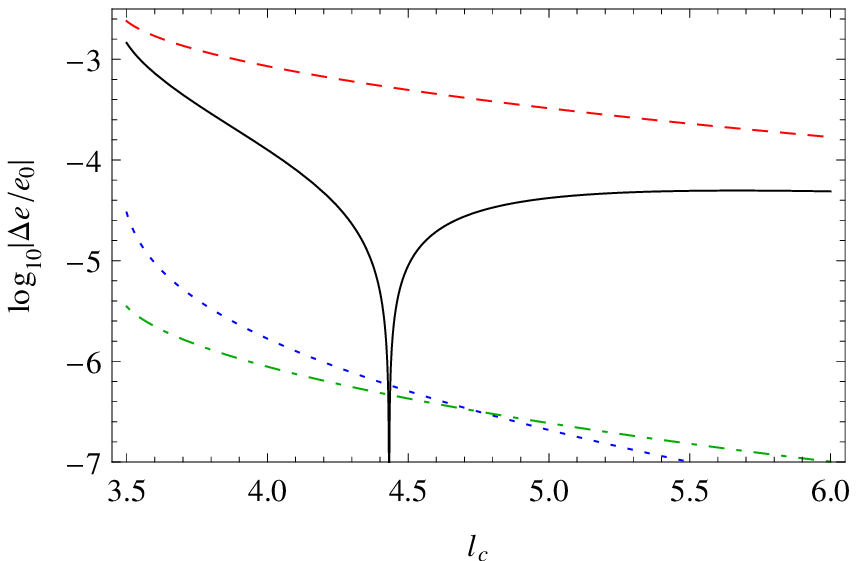}\label{spinPlotBH}}
\caption{Plots illustrating spin and self-force corrections to the binding energy of a multipolar object in Schwarzschild spacetime $\hat{a}_1 = 0$. In \subref{spinPlot} and \subref{spinPlotZoom} the corrections are normalized by subtracting their dependence on $q$, $\hat{a}_2$, and $C_{ES^2}$, see (\ref{SFcurve}), (\ref{Scurve}), (\ref{S2curve}), and (\ref{CEScurve}) [belonging to curves (*), (**), (***), and (****), respectively].\label{spinPlots}}
\end{center}
\end{figure}

In order to better separate the different contributions to the binding energy, we make a series expansion of the form
\begin{equation}
\hat{r} = \hat{r}_0 + \epsilon \, \hat{r}_1 + \epsilon^2 \, \hat{r}_2 + \dots.
\end{equation}
This, together with $\hat{J}=l_c+q \hat{a}_2$, is inserted into (\ref{circ_cond}) and the whole expression is expanded in $\epsilon$. Then one can solve for $\hat{r}$ order-by-order in $\epsilon$, i.e., the $\epsilon^0$-part of (\ref{circ_cond}) is solved for $\hat{r}_0$, the $\epsilon^1$-part of (\ref{circ_cond}) is solved for $\hat{r}_1$, and so on. The solution for $\hat{r}$ is then plugged into $e(\hat{r}, l_c)$ leading to $e(l_c)$, which is expanded again,
\begin{equation}
e(l_c) = e_0(l_c) + \epsilon \, e_1(l_c) + \epsilon^2 \, e_2(l_c) + \dots .
\end{equation}
The correction $e_1(l_c)$ does not contain quadrupole contributions and is linear in the spin $\hat{a}_2$ and mass ratio $q$. In the figs.\ \ref{spinPlot} and \ref{spinPlotZoom} we have plotted the ``normalized'' relative difference to the leading order $e_0(l_c)$,
\begin{equation}
\log_{10} \left| \frac{e_1(l_c)}{e_0(l_c)} \right| - \log_{10} \left| q \hat{a}_2 \right|, \label{Scurve}
\end{equation}
which does not depend on $\hat{a}_2$ and $q$ (this corresponds to setting $\hat{a}_2 = 1 = q$). The correction $e_2(l_c)$ can be split as
\begin{equation}
e_2(l_c) = e_2^{S^2}(l_c) + e_2^{C_{ES^2}}(l_c) + e_2^{k_2}(l_c) + e_2^{j_2}(l_c) ,
\end{equation}
into tidal quadrupole contributions $e_2^{k_2}(l_c)$ and $e_2^{j_2}(l_c)$ proportional to $k_2$ and $j_2$, respectively, a quadratic-in-spin quadrupole $e_2^{C_{ES^2}}(l_c)$ proportional to $C_{ES^2}$, and a remaining term $e_2^{S^2}(l_c)$ which is quadratic in spin. 

The spin parts are again normalized,
\begin{eqnarray}
&& \log_{10} \left| \frac{e_2^{S^2}(l_c)}{e_0(l_c)} \right| - \log_{10} \left| q^2 \hat{a}_2^2 \right| , \label{S2curve} \\
&& \log_{10} \left| \frac{e_2^{C_{ES^2}}(l_c)}{e_0(l_c)} \right| - \log_{10} \left| C_{ES^2} q^2 \hat{a}_2^2 \right| , \label{CEScurve}
\end{eqnarray}
and plotted in figs.\ \ref{spinPlot} and \ref{spinPlotZoom} -- tidal effects will be discussed in the next section. These plots also show recent results for the conservative part of the self-force normalized as
\begin{equation}
\log_{10} \left| \frac{e_{\text{TBB}}(l_c)-e_0(l_c)}{e_0(l_c)} \right| - \log_{10} \left| q \right|, \label{SFcurve}
\end{equation}
where $e_{\text{TBB}}(l_c)$ is the binding energy given in a parametric form in \cite{LeTiec:Barausse:Buonanno:2012}, see also app.\ \ref{sec_ca_results} for an explicit expression. The difference $e_{\text{TBB}}(l_c)-e_0(l_c)$ changes its sign at a certain value of $l_c$, which appears in the plots as a pole. It should be stressed that our results do not include contributions that arise from a perturbation of the background spacetime due to the spin or quadrupole of the test body.

Notice that the transition from the normalized curves to realistic curves in figs.\ \ref{spinPlot} and \ref{spinPlotZoom} is straigtforward. If one considers a mass ratio of $q = 10^{-2}$, then the self-force curve (\ref{SFcurve}) is just shifted down by $\log_{10} q = -2$. Similarly, for a rotating star -- with $\hat{a}_2 = 10^{-0.5} \approx 0.32$ and $C_{ES^2} = 10^{0.5} \approx 3.2$ -- the correction linear in spin (\ref{Scurve}) is shifted down by $-2.5$, the curve (\ref{S2curve}) is shifted down by $-5$, and the $C_{ES^2}$-quadrupole curve (\ref{CEScurve}) is shifted down by $-4.5$ (still for $q = 10^{-2}$). For a black hole one may even assume $\hat{a}_2 \sim 1$, furthermore we have $C_{ES^2}= 1$. The latter case is illustrated in fig.\ \ref{spinPlotBH}. Notice that the dominant scaling is coming from the mass ratio $q$. The linear-in-spin correction and the self-force scale with the same power of $q$. One can also easily obtain the sign of the correction $\Delta e$ to the binding energy, it holds
\begin{eqnarray}
&& \text{sgn}\left(e_1\right) = \text{sgn}\left(\hat{a}_2\right) , \quad \text{sgn}\left(e_2^{S^2}\right) = - 1 , \\
&& \text{sgn}\left(e_2^{C_{ES^2}}\right) = - \text{sgn}\left(C_{ES^2}\right) ,
\end{eqnarray}
while the self-force correction $e_{\text{TBB}}-e_0$ has a positive sign for small $l_c$ and a negative sign for large $l_c$.

The plots confirm that the multipole approximation introduced in the present paper is well justified, as the quadrupole effects are always at least an order of magnitude weaker than the linear spin effects -- within the context of the quadrupole model adopted in this work. Also notice that the quadrupole effects scale with a higher power of the mass ratio $q$ than the linear spin corrections.

\begin{figure}
\begin{center}
\includegraphics{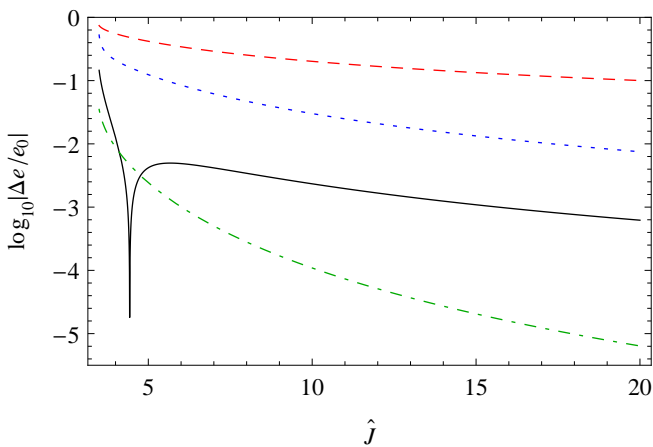}
\end{center}
\caption{\label{fig_energy_ito_j} Binding energy in terms of $\hat{J}$. This plot makes clear that one has to be careful when in comes to the parameterization of the strength of different corrections. In contrast to the $l_c$-parameterization, effects from the internal structure -- in this case the spin -- of the test body, appear more pronounced in the $\hat{J}$-parametrization. The quadrupole curve is not affected.}
\end{figure}

To further illustrate the difference between the parameters $l_c$ and $\hat{J}$, we also include a plot of the binding energy $e(\hat{J})$ in fig.\ \ref{fig_energy_ito_j}.

\subsubsection{Tidal Effects}

\begin{figure}
\begin{center}
\subfigure[$\quad$Normalized curves for tidal and self-force corrections.]
{\includegraphics{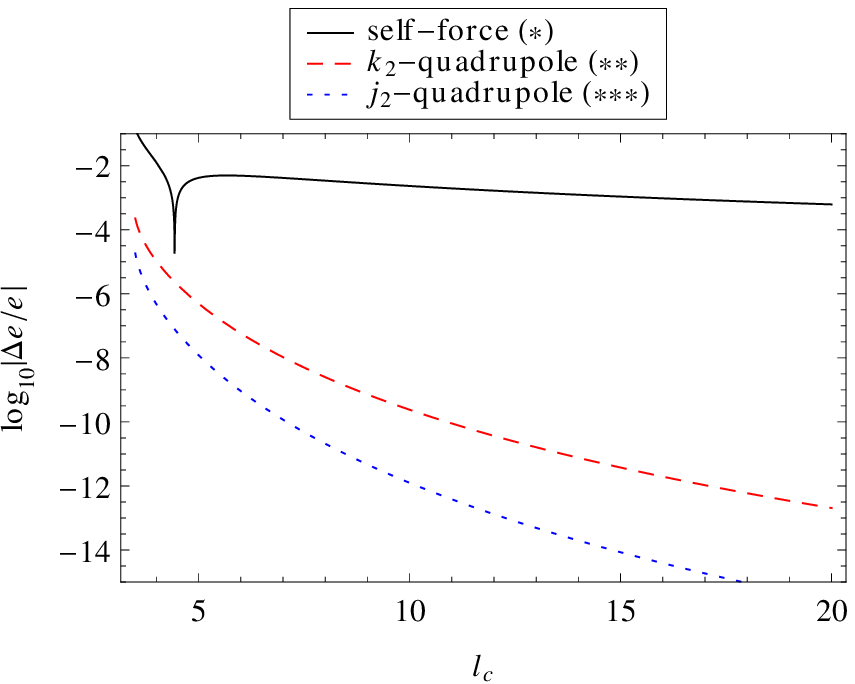}\label{tidePlot}}
\subfigure[$\quad$The same as \subref{tidePlot}, but zoomed in at small $l_c$.]
{\includegraphics{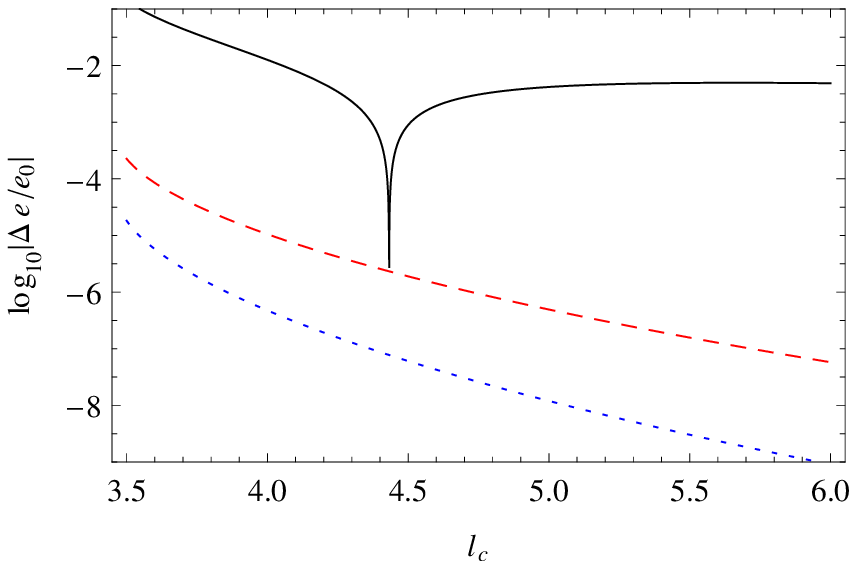}\label{tidePlotZoom}}
\subfigure[$\quad$Similar as above, but the normalized curves are shifted down to a mass ratio $q=10^{-4}$ for $k_2 = 0.01$ and $\hat{R} = 10^4$, which can occur for a white dwarf. Tidal disruption can be expected when the $k_2$-curve reaches $-6.6$.]
{\includegraphics{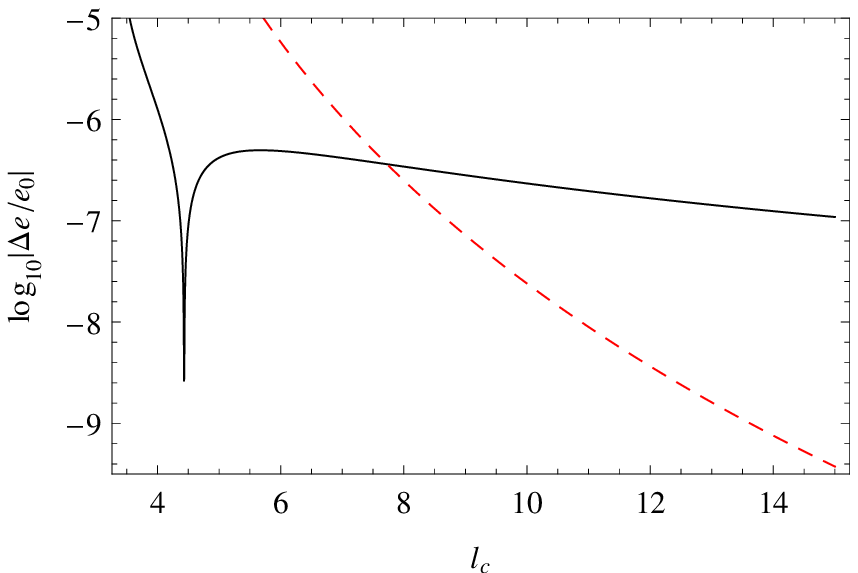}\label{tidePlotWD}}
\caption{Plots illustrating tidal and self-force corrections to the binding energy of a multipolar object in Schwarzschild spacetime $\hat{a}_1 = 0$. In \subref{tidePlot} and \subref{tidePlotZoom} the corrections are normalized by subtracting their dependence on $q$, $\hat{R}$, $k_2$, and $j_2$, see (\ref{SFcurve}), (\ref{k2curve}), and (\ref{j2curve}) [belonging to curves (*), (**), and (***), respectively].\label{tidePlots}}
\end{center}
\end{figure}

For the tidal contributions we also plot normalized quantities,
\begin{eqnarray}
&&\log_{10} \left| \frac{e_2^{k_2}(l_c)}{e_0(l_c)} \right| - \log_{10} \left| k_2 q^4 \hat{R}^5 \right| , \label{k2curve} \\
&&\log_{10} \left| \frac{e_2^{j_2}(l_c)}{e_0(l_c)} \right| - \log_{10} \left| j_2 q^4 \hat{R}^5 \right| , \label{j2curve}
\end{eqnarray}
see figs.\ \ref{tidePlot} and \ref{tidePlotZoom}. Notice that the change in the effective potential due to tidal deformations, and thus their backreaction to the orbit, is proportional to the difference between dynamical mass $\underline{m}$ and constant mass parameter $\mu$. The scaling in $q$ and $\hat{R}$ then immediately follows from (\ref{mquad}). This suggests that tidal effects are most relevant for big (non-compact) objects due to the proportionality to $\hat{R}^5$. However, due to tidal disruption the maximal
possible tidal backreaction is proportional to $1/\hat{R}$, cf.\ eq.\ (\ref{disrupt}). Therefore, in order to produce a strong tidal backreaction, it is in fact important to consider objects which are very compact (like white dwarfs or neutron stars), such that their own gravitational field can still hold the object together when the tidal forces become strong.

For example, for a white dwarf one can take $k_2 = 0.01$ and $\hat{R} = 10^4$. For a mass ratio of $q = 10^{-4}$ the $k_2$-quadrupole curve is then shifted \emph{up} by $+2$, while the self-force curve is shifted down by $-4$. Tidal disruption can be expected when the $k_2$-curve reaches
\begin{equation}
\log_{10} \left( \frac{k_2}{4 \hat{R}} \right) \approx -6.6 ,
\end{equation}
see (\ref{disrupt}). Thus, the tidal backreaction can reach the same order as the self-force before the object is disrupted. For neutron stars one has $\hat{R} \sim 5$ and $k_2 = 0.1$, so disruption only happens at about $-2.3$. However, this Newtonian estimate might not be very accurate for neutron stars. On the other hand, the $k_2$-curve for a neutron star is shifted further \emph{down} by $-1.5$ even for a mass ratio $q = 10^{-1}$ (where the used approximation is not accurate any more). For neutron stars the tidal backreaction is most relevant for comparable masses, as it can become very strong in this case. The signs of the corrections are given by
\begin{equation}
\text{sgn}\left(e_2^{k_2}\right) = - \text{sgn}\left(k_2\right) , \quad
\text{sgn}\left(e_2^{j_2}\right) = - \text{sgn}\left(j_2\right) .
\end{equation}

\subsection{Corrections in Kerr spacetime\label{sec_corr_in_Kerr}}

\begin{figure}
\begin{center}
\subfigure[\quad Linear in spin $\sim \hat{a}_2$, eq.\ (\ref{Scurve})]
{\includegraphics{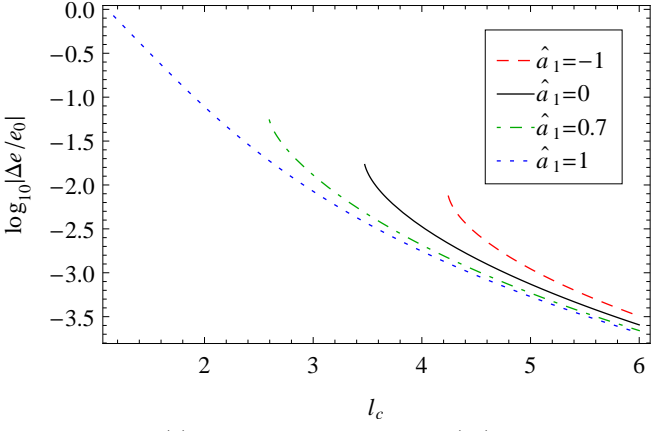}\label{KerrPlotE1}}
\subfigure[\quad Quadratic in spin $\sim \hat{a}_2^2$, eq.\ (\ref{S2curve})]
{\includegraphics{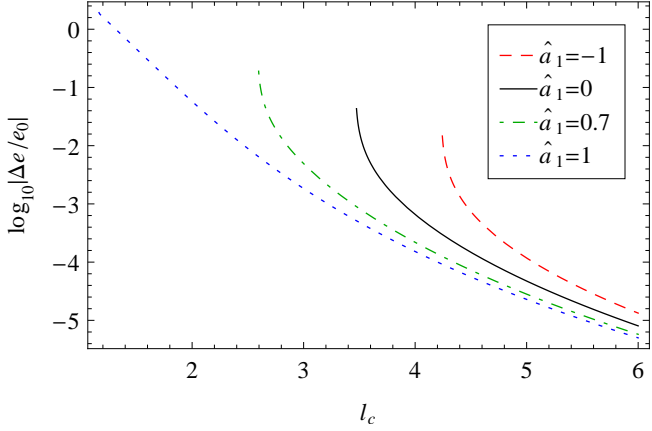}\label{KerrPlotS2}}
\caption{Pole-Dipole corrections in Kerr spacetime.\label{KerrPoltsS}}
\end{center}
\end{figure}

\begin{figure}
\begin{center}
\subfigure[\quad $C_{ES^2}$-quadrupole proportional to $\hat{a}_2^2$, eq.\ (\ref{CEScurve})]
{\includegraphics{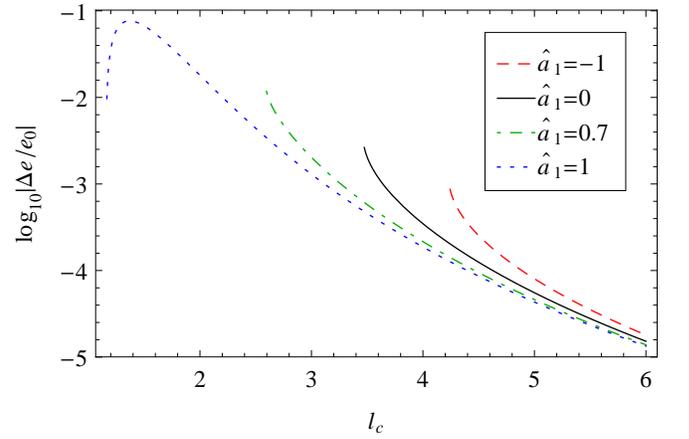}\label{KerrPlotCES}}
\subfigure[\quad  Tidal $k_2$-quadrupole, eq.\ (\ref{k2curve})]
{\includegraphics{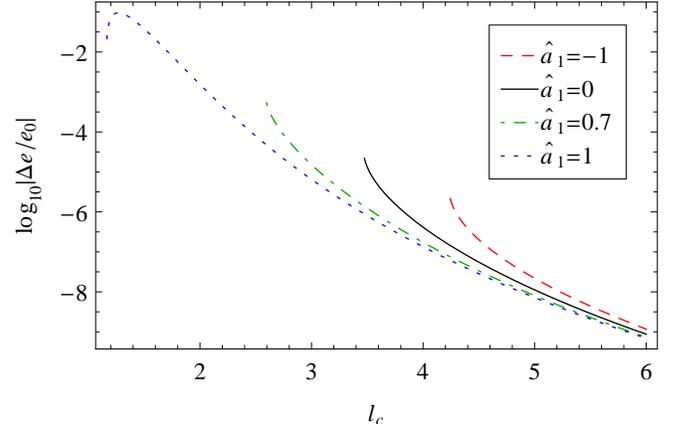}\label{KerrPlotK2}} \subfigure[\quad  Tidal $j_2$-quadrupole, eq.\ (\ref{j2curve})]
{\includegraphics{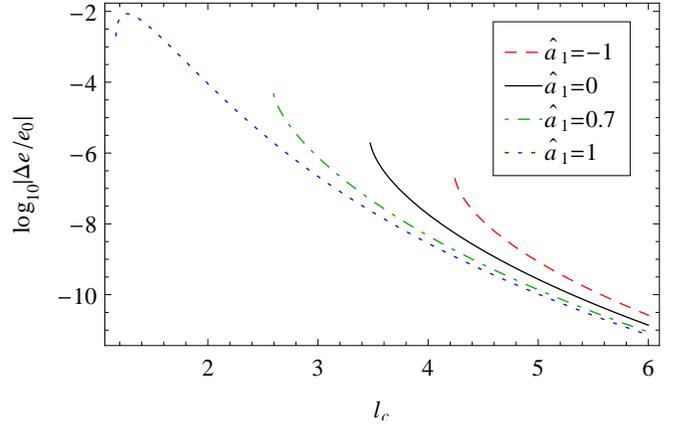}\label{KerrPlotJ2}}
\caption{Quadrupole corrections in Kerr spacetime.\label{KerrPoltsQ}}
\end{center}
\end{figure} 

For Kerr spacetime, $\hat{a}_1 \neq 0$, one can still separate the various contributions to the binding energy. This is completely analogous to the case $\hat{a}_1 = 0$, except that we are not solving for $r_0$ analytically. Still one can write the zeroth-order approximation of (\ref{circ_cond}) as \cite[eq.\ (2.13)]{Bardeen:Press:Teukolsky:1972}
\begin{equation}
l_c = \frac{\hat{r}_0^2 - 2 \hat{a}_1 \sqrt{\hat{r}_0} + \hat{a}_1^2}{\sqrt{\hat{r}_0} \sqrt{\hat{r}_0^2 + 2 \hat{a}_1 \sqrt{\hat{r}_0} - 3 \hat{r}_0}} ,
\end{equation}
which is most conveniently inverted numerically.

Plots for the linear and quadratic spin contributions for a pole-dipole particle are shown in fig.\ \ref{KerrPoltsS}, and quadrupole contributions can be found in fig.\ \ref{KerrPoltsQ}. For negative $\hat{a}_1$ the contributions are stronger than the Schwarzschild ones for the same value of $l_c$. However, the last stable circular orbit is also reached at greater values of $l_c$, so finally larger corrections are possible in the Schwarzschild case (at values of $l_c$ where the Kerr orbits are already unstable). For positive $\hat{a}_1$ the opposite is true: the curve is below the Schwarzschild one, but the last stable circular orbit is located at lower $l_c$. For the case $\hat{a}_1 = 1$ the last stable orbit reaches the horizon of the black hole. Therefore the test body can orbit much closer to the horizon, where the field is very strong. This finally allows the same contributions to become one or two -- or even more for tidal effects -- orders of magnitude stronger than in the Schwarzschild case if $\hat{a}_1 >0$. This can make tidal effects for neutron stars interesting again even beyond the comparable mass case.

\subsection{Comparison with post-Newtonian Hamiltonians\label{sec:PN}}

We will now compare our result for the binding energy $e(\hat{J})$ with post-Newtonian Hamiltonians. We first calculate the post-Newtonian expansion of our result for $e(\hat{J})$ (which is valid to all post-Newtonian orders). Then again we calculate $e(\hat{J})$ from post-Newtonian Hamiltonians (valid for generic mass ratio $q$) and expand in the mass ratio $q$. Both should lead to the same result, as the regimes of approximation overlap and the relation $e(\hat{J})$ is gauge invariant. We include in our comparison all (conservative) post-Newtonian Hamiltonians in the Arnowitt-Deser-Misner transverse-traceless (ADMTT) gauge known to date.

We start with a definition of the post-Newtonian expansion. The post-Newtonian approximation is a weak field ($\hat{r} \gg 1$) and slow motion ($l_c \gg 1$ or $\hat{J} \gg 1$) approximation. For bound orbits the Newtonian virial theorem establishes a relation $\hat{J}^2 \sim \hat{r}$. It is convenient to introduce a post-Newtonian book-keeping parameter $\epsilon_{PN}$ and associate power counting rules to all variables,
\begin{equation}
l_c = \Order(\epsilon_{PN}^{-1}), \quad \hat{J} = \Order(\epsilon_{PN}^{-1}), \quad \hat{r} = \Order(\epsilon_{PN}^{-2}).
\end{equation}
Notice that $\hat{a}_1$, $\hat{a}_2$, $C_{ES^2}$, $k_2$, $j_2$, and $\hat{R}$ are in fact further expansion variables that are not related to the post-Newtonian approximation, which suggests to count them as $\Order(\epsilon_{PN}^0)$. This approach will indeed be followed here with the only exception that we formally assume
\begin{equation}
\hat{R} = \Order(\epsilon_{PN}^{-1}) .
\end{equation}
This choice is of course quite arbitrary and is only made because otherwise the tidal interactions would formally appear at very high post-Newtonian order only (which somewhat underestimates their effect and would require us to expand to very high order for the comparison).

Using these power counting rules it is straightforward to expand $e(\hat{r}, \hat{J})$ in $\epsilon_{PN}$,
\begin{widetext}
\begin{eqnarray}
e(\hat{r}, \hat{J}) &=& \frac{1}{\hat{r}} \bigg[ -1	 + \frac{\hat{J}^2}{2 \hat{r}} \bigg] - \frac{\hat{a}_2 \hat{J} q}{\hat{r}^2} + \frac{1}{\hat{r}^2} \bigg[-\frac{1}{2} - \frac{\hat{J}^4}{8 \hat{r}^2} - \frac{\hat{J}^2}{2 \hat{r}} + \frac{\hat{a}_2^2 q^2}{2} \bigg] \nonumber\\ 
&& + \frac{\hat{J}}{\hat{r}^3} \bigg[ 2 \hat{a}_1 + 3 \hat{a}_2 q \bigg] + \frac{1}{\hat{r}^3} \bigg[-\frac{1}{2} - \frac{\hat{J}^2}{4 \hat{r}} + \frac{\hat{J}^4}{8 \hat{r}^2} + \frac{\hat{J}^6}{16 \hat{r}^3} - \frac{\hat{a}_1^2 \hat{J}^2}{2 \hat{r}} - 3 \hat{a}_1 \hat{a}_2 q - \frac{\hat{a}_2^2 q^2}{2} (C_{ES^2}+5) + \frac{3 \hat{a}_2^2 \hat{J}^2 q^2}{4 \hat{r}} \bigg] \nonumber\\ 
&& + \frac{\hat{J}}{\hat{r}^4} \bigg[ \hat{a}_1^2 \hat{a}_2 q -  \frac{k_2 q^4 \hat{R}^5}{\hat{J} \hat{r}^2} \bigg] + \frac{1}{\hat{r}^4} \bigg[ -\frac{5}{8} - \frac{\hat{J}^2}{4 \hat{r}} + \frac{\hat{J}^4}{16 \hat{r}^2} - \frac{\hat{J}^6}{16 \hat{r}^3} - \frac{5 \hat{J}^8}{128 \hat{r}^4} + 2 \hat{a}_1^2 + \frac{\hat{a}_1^2 \hat{J}^4}{4 \hat{r}^2} - \frac{\hat{a}_1^2 \hat{J}^2}{2 \hat{r}} - \frac{\hat{a}_1^2 \hat{a}_2^2 q^2}{2} \nonumber \\ 
&& \qquad + 3 \hat{a}_1 \hat{a}_2 q - \frac{9 \hat{a}_1 \hat{a}_2 \hat{J}^2 q}{2 \hat{r}} + \frac{\hat{a}_2^2 q^2}{4} (2C_{ES^2}+7) - \frac{\hat{a}_2^2 \hat{J}^2 q^2}{4 \hat{r}} (5C_{ES^2}+11) - \frac{5\hat{a}_2^2 \hat{J}^4 q^2}{16 \hat{r}^2} \bigg] \nonumber\\ 
&& + \frac{\hat{J}}{\hat{r}^5} \bigg[ -2 \hat{a}_1^3 + \hat{a}_1^2 \hat{a}_2 q + 3 \hat{a}_1 \hat{a}_2^2 q^2 (C_{ES^2}+3) + \frac{k_2 q^4 \hat{R}^5}{\hat{J} \hat{r}^2} - \frac{\hat{J} q^4 \hat{R}^5}{4 \hat{r}^3} (10k_2+j_2) \bigg] \nonumber\\ 
&& + \frac{1}{\hat{r}^5} \bigg[ -\frac{7}{8} - \frac{5 \hat{J}^2}{16 \hat{r}} + \frac{\hat{J}^4}{16 \hat{r}^2} - \frac{\hat{J}^6}{32 \hat{r}^3} + \frac{5 \hat{J}^8}{128 \hat{r}^4} + \frac{7 \hat{J}^{10}}{256 \hat{r}^5} + \frac{\hat{a}_1^4 \hat{J}^2}{2 \hat{r}} + 2 \hat{a}_1^3 \hat{a}_2 q + 2 \hat{a}_1^2 + \frac{9 \hat{a}_1^2 \hat{J}^2}{4 \hat{r}} + \frac{\hat{a}_1^2 \hat{J}^4}{4 \hat{r}^2} - \frac{3 \hat{a}_1^2 \hat{J}^6}{16 \hat{r}^3} \nonumber\\ 
&& \qquad - \frac{\hat{a}_1^2 \hat{a}_2^2 q^2}{2} (3C_{ES^2}+1) - \frac{3 \hat{a}_1^2 \hat{a}_2^2 \hat{J}^2 q^2}{2 \hat{r}} + \frac{3 \hat{a}_1 \hat{a}_2 q}{2} + \frac{9 \hat{a}_1 \hat{a}_2 \hat{J}^2 q}{2 \hat{r}} + \frac{15 \hat{a}_1 \hat{a}_2 \hat{J}^4 q}{8 \hat{r}^2} + \frac{\hat{a}_2^2 q^2}{4} (C_{ES^2}+3) \nonumber\\ 
&& \qquad + \frac{\hat{a}_2^2 \hat{J}^2 q^2}{8 \hat{r}} (10C_{ES^2}+13) + \frac{\hat{a}_2^2 \hat{J}^4 q^2}{16 \hat{r}^2} (9C_{ES^2}+17) + \frac{7 \hat{a}_2^2 \hat{J}^6 q^2}{32 \hat{r}^3} + \frac{\hat{a}_1 \hat{J} q^4 \hat{R}^5}{2\hat{r}^3} (12k_2+j_2) \bigg] \nonumber\\ 
&& + \frac{1}{\hat{r}^6} \bigg[-  \hat{a}_1^4 \hat{a}_2 q	 -4 \hat{a}_1^3 - 18 \hat{a}_1^2 \hat{a}_2 q - 3 \hat{a}_1 \hat{a}_2^2 q^2 (3C_{ES^2}+7) + \frac{k_2 q^4 \hat{R}^5}{2 \hat{J} \hat{r}^2} - \frac{\hat{a}_1^2 q^4 \hat{R}^5}{4 \hat{J} \hat{r}^2} (12k_2+j_2) \nonumber\\ 
&& \qquad + \frac{\hat{J} q^4 \hat{R}^5}{4 \hat{r}^3} (10k_2+j_2) - \frac{\hat{J}^3 q^4 \hat{R}^5}{8 \hat{r}^4} (15k_2+j_2) \bigg] + \Order(\epsilon^3, \epsilon_{PN}^{12}) . \label{erj_PN}
\end{eqnarray}
Next we solve the condition defining circular orbits (\ref{circ_cond}) for $\hat{r}$ (order-by-order in $\epsilon_{PN}$) and insert the result into (\ref{erj_PN}). This leads to the post-Newtonian expanded gauge-invariant relation
\begin{eqnarray}
e(\hat{J}) &=& -\frac{1}{2 \hat{J}^2} -  \frac{\hat{a}_2 q}{\hat{J}^3} + \frac{1}{\hat{J}^4} \bigg[ -\frac{9}{8} - \frac{3}{2} \hat{a}_2^2 q^2 \bigg] + \frac{1}{\hat{J}^5} \bigg[ 2 \hat{a}_1 - 3 \hat{a}_2 q \bigg] + \frac{1}{\hat{J}^6} \bigg[ -\frac{81}{16} - \frac{1}{2} \hat{a}_1^2 + 9 \hat{a}_1 \hat{a}_2 q - \frac{1}{4} \hat{a}_2^2 q^2 (2C_{ES^2}+15) \bigg] \nonumber\\ 
&& + \frac{1}{\hat{J}^7} \bigg[18 \hat{a}_1 - 18 \hat{a}_2 q - 3 \hat{a}_1^2 \hat{a}_2 q + 24 \hat{a}_1 \hat{a}_2^2 q^2 - k_2 q^4 \frac{\hat{R}^5}{\hat{J}^5} \bigg] \nonumber\\ 
&& + \frac{1}{\hat{J}^8} \bigg[ -\frac{3861}{128} - \frac{89}{4} \hat{a}_1^2 + \frac{183}{2} \hat{a}_1 \hat{a}_2 q - \frac{21}{2} \hat{a}_1^2 \hat{a}_2^2 q^2 - \frac{21}{16} \hat{a}_2^2 q^2 ( 4 C_{ES^2} + 21 ) \bigg] \nonumber\\ 
&& + \frac{1}{\hat{J}^9} \bigg[ 162 \hat{a}_1 + 10 \hat{a}_1^3 - 135 \hat{a}_2 q - 151 \hat{a}_1^2 \hat{a}_2 q + 12 \hat{a}_1 \hat{a}_2^2 q^2 (C_{ES^2}+20) - \frac{q^4 \hat{R}^5}{4 \hat{J}^5} (78k_2+j_2) \bigg] \nonumber\\ 
&& + \frac{1}{\hat{J}^{10}} \bigg[ -\frac{53703}{256} - \frac{5643}{16} \hat{a}_1^2 - \frac{3}{2} \hat{a}_1^4 + \frac{7623}{8} \hat{a}_1 \hat{a}_2 q + 84 \hat{a}_1^3 \hat{a}_2 q \nonumber\\ 
&& \qquad - \frac{9}{32} \hat{a}_2^2 q^2 (182 C_{ES^2} + 907) - \frac{9}{2} \hat{a}_1^2 \hat{a}_2^2 q^2 (C_{ES^2}+125) + \frac{\hat{a}_1 q^4 \hat{R}^5}{2\hat{J}^5} (84k_2+j_2) \bigg] \nonumber\\ 
&& + \frac{1}{\hat{J}^{11}} \bigg[ 1512 \hat{a}_1 + 386 \hat{a}_1^3 - 1134 \hat{a}_2 q - 2619 \hat{a}_1^2 \hat{a}_2 q - 15 \hat{a}_1^4 \hat{a}_2 q + 6 \hat{a}_1 \hat{a}_2^2 q^2 (37 C_{ES^2} + 453) + 390 \hat{a}_1^3 \hat{a}_2^2 q^2 \nonumber\\ 
&& \qquad - \frac{q^4 \hat{R}^5}{\hat{J}^5} \bigg(\frac{2247}{8} k_2	+ 15 \hat{a}_1^2 k_2	+ \frac{47}{8} j_2+ \frac{1}{4} \hat{a}_1^2 j_2 \bigg)\bigg] + \Order(\epsilon^3, \epsilon_{PN}^{12}) . \label{ejPN}
\end{eqnarray}
\end{widetext}

Now we establish the connection to the (conservative) post-Newtonian Hamiltonian $H$ within the ADMTT gauge. We include the Hamiltonians
\begin{eqnarray}
H &=& H_{\text{N}} + H_{\text{1PN}} + H_{\text{2PN}} + H_{\text{3PN}}+ H_{\text{LO}}^{\text{SO}} + H_{\text{NLO}}^{\text{SO}} \nonumber \\ 
&&	+ H_{\text{NNLO}}^{\text{SO}}	+ H_{\text{LO}}^{S_1S_2} + H_{\text{NLO}}^{S_1S_2} + H_{\text{NNLO}}^{S_1S_2}	+ H_{\text{LO}}^{S^2} \nonumber  \\ 
&&	+ H_{\text{NLO}}^{S^2}+ H_{\text{LO}}^{S^3}	+ H^{k_2}_{\text{LO}} + H^{j_2}_{\text{LO}}	+ \dots ,
\end{eqnarray}
which we will list in the following in the center-of-mass frame for aligned spins together with the corresponding literature. Further, we already expand the Hamiltonians in the mass ratio $q$ to the order needed for the comparison.

The Newtonian (N), first post-Newtonian (1PN), and second post-Newtonian (2PN)
Hamiltonians are given by
\begin{eqnarray}
H_{\text{N}} &=& - \frac{1}{\hat{r}_{c}} \bigg[ 1-\frac{l_c^2}{2 \hat{r}_{c}} \bigg] , \\
H_{\text{1PN}} &=& \frac{1}{2 \hat{r}_{c}^2} \bigg[ 1-\frac{3 l_c^2}{\hat{r}_{c}}-\frac{l_c^4}{4 \hat{r}_{c}^2} \bigg] , \\
H_{\text{2PN}} &=& - \frac{1}{4 \hat{r}_{c}^3} \bigg[ 1-\frac{10 l_c^2}{\hat{r}_{c}}-\frac{5 l_c^4}{2 \hat{r}_{c}^2}-\frac{l_c^6}{4 \hat{r}_{c}^3} \bigg], 
\end{eqnarray}
see, e.g., \cite{Ohta:Okamura:Kimura:Hiida:1974, Damour:Schafer:1985,Damour:Schafer:1988}, and references therein. Here $\hat{r}_{c}$ denotes the ADMTT-gauge (canonical) radial coordinate. The third post-Newtonian level was first tackled in \cite{Jaranowski:Schafer:1998}. But the result contained two ``ambiguity'' parameters, which were subsequently determined \cite{Jaranowski:Schafer:1999,Damour:Jaranowski:Schafer:2000}, see also \cite{Kimura:Toiya:1972}. (Dimensional regularization must be used to avoid such ambiguities \cite{Damour:Jaranowski:Schafer:2001}.) An alternative derivation of the equations of motion at the third post-Newtonian order can also be found in \cite{Futamase:Itoh:2003, Blanchet:Damour:EspositoFarese:2004}. The full result in the test body limit reads
\begin{equation}
H_{\text{3PN}} = \frac{1}{8 \hat{r}_{c}^4} \bigg[ 1-\frac{25 l_c^2}{\hat{r}_{c}}-\frac{27 l_c^4}{2 \hat{r}_{c}^2}-\frac{7 l_c^6}{2 \hat{r}_{c}^3}-\frac{5 l_c^8}{16 \hat{r}_{c}^4} \bigg] .
\end{equation}
The leading order (LO) spin-orbit (SO) and $S_1S_2$ Hamiltonians are given by
\begin{eqnarray}
H_{\text{LO}}^{\text{SO}} &=&\frac{2 \hat{a}_1 l_c}{\hat{r}_{c}^3}+\frac{3 q \hat{a}_2 l_c}{2 \hat{r}_{c}^3} , \\ 
H_{\text{LO}}^{S_1S_2} &=& -\frac{q \hat{a}_1 \hat{a}_2}{\hat{r}_{c}^3} ,
\end{eqnarray}
see \cite{Barker:OConnell:1975,DEath:1975,Barker:OConnell:1979,Thorne:Hartle:1985}. These references also contain the $S_1S_1$ interaction potential for the black hole case $C_{ES^2} = 1$. The parameter $C_{ES^2}$ was introduced in \cite{Poisson:1998}. The sum of $S_1S_1$ and $S_2S_2$ interactions can be written as
\begin{equation}
H_{\text{LO}}^{S^2} = -\frac{\hat{a}_1^2}{2 \hat{r}_{c}^3}-\frac{C_{ES^2} q^2 \hat{a}_2^2}{2 \hat{r}_{c}^3} .
\end{equation}
The next-to-leading order (NLO) spin-orbit Hamiltonian was derived within the ADMTT gauge in \cite{Damour:Jaranowski:Schafer:2008:1,
Steinhoff:Schafer:Hergt:2008},
\begin{equation}
H_{\text{NLO}}^{\text{SO}} = -\frac{6 \hat{a}_1 l_c}{\hat{r}_{c}^4} - \frac{5 q \hat{a}_2 l_c}{\hat{r}_{c}^4} \bigg[ 1+\frac{l_c^2}{8 \hat{r}_{c}} \bigg] .
\end{equation}
Other results on this interaction can be found in \cite{Tagoshi:Ohashi:Owen:2001,Faye:Blanchet:Buonanno:2006,Levi:2010,Porto:2010,Perrodin:2010,Hergt:Steinhoff:Schafer:2011}. The next-to-leading order $S_1S_2$ Hamiltonian \cite{Steinhoff:Hergt:Schafer:2008:2,Steinhoff:Schafer:Hergt:2008} is given by
\begin{equation}
H_{\text{NLO}}^{S_1S_2} = \frac{6 q \hat{a}_1 \hat{a}_2}{\hat{r}_{c}^4} \bigg[ 1-\frac{l_c^2}{4 \hat{r}_{c}} \bigg] , 
\end{equation}
see also \cite{Porto:Rothstein:2008:1, Levi:2008, Hergt:Steinhoff:Schafer:2011}. The next-to-leading order $S_1S_1$ and $S_2S_2$ interaction Hamiltonians reads \cite{Steinhoff:Hergt:Schafer:2008:1, Hergt:Steinhoff:Schafer:2010:1}
\begin{eqnarray}
H_{\text{NLO}}^{S^2} &=& \frac{q^2 \hat{a}_2^2}{\hat{r}_{c}^4} \bigg[ (2 C_{ES^2}+1)-\frac{5 (C_{ES^2}-1) l_c^2}{4 \hat{r}_{c}} \bigg] \nonumber \\
&&	+ \frac{\hat{a}_1^2}{2 \hat{r}_{c}^4} \bigg[ 5-\frac{3 l_c^2}{2 \hat{r}_{c}} \bigg] ,
\end{eqnarray}
see also \cite{Porto:Rothstein:2008:2, Steinhoff:Schafer:2009:1,Porto:Rothstein:2008:2:err,Hergt:Steinhoff:Schafer:2011}. Recently even
next-to-next-to-leading order (NNLO) spin interaction Hamiltonians were calculated, namely the spin-orbit one in the test-spin limit \cite{Barausse:Racine:Buonanno:2009} (see \cite{Hartung:Steinhoff:2011:1} for the complete result)
\begin{equation}
H_{\text{NNLO}}^{\text{SO}} = \frac{21 \hat{a}_1 l_c}{2 \hat{r}_{c}^5} + \frac{3 q \hat{a}_2 l_c}{8 \hat{r}_{c}^5} \bigg[ 25+\frac{9 l_c^2}{\hat{r}_{c}}+\frac{7 l_c^4}{6 \hat{r}_{c}^2} \bigg] ,
\end{equation}
and also the $S_1S_2$ Hamiltonian \cite{Hartung:Steinhoff:2011:2,Levi:2011}
\begin{equation}
H_{\text{NNLO}}^{S_1S_2} = -\frac{9 q \hat{a}_1 \hat{a}_2}{4 \hat{r}_{c}^5} \bigg[ 7-\frac{4 l_c^2}{\hat{r}_{c}}-\frac{l_c^4}{2 \hat{r}_{c}^2} \bigg] .
\end{equation}
The Hamiltonians cubic in the spins derived in \cite{Hergt:Schafer:2008:2,Hergt:Schafer:2008} are only valid for binary black holes (i.e., $C_{ES^2}$ = 1) and can be summarized as
\begin{equation}
H_{\text{LO}}^{S^3} = \frac{l_c}{4 \hat{r}_{c}^5} (\hat{a}_1+q \hat{a}_2)^2 (4 \hat{a}_1+q \hat{a}_2) ,
\end{equation}
see also \cite{Barausse:Racine:Buonanno:2009}.\footnote{Also Hamiltonians of quartic order in spin are given in \cite{Hergt:Schafer:2008:2, Hergt:Schafer:2008}. However, complete agreement with the results of the present paper could not be found yet. The deviation indicates that the $S^4$ Hamiltonian discussed in \cite {Hergt:Schafer:2008} does not vanish, in contrast to the conclusion therein. Also notice that a misprint at quartic order in spin was corrected in the arXiv version of \cite{Hergt:Schafer:2008:2} recently.} The leading order tidal interaction Hamiltonians read \cite{Damour:Nagar:2009}
\begin{equation}
H^{k_2}_{\text{LO}} = - \frac{k_2 q^4 \hat{R}^5}{\hat{r}_{c}^6} , \qquad H^{j_2}_{\text{LO}} = - \frac{j_2 q^4 \hat{R}^5 l_c^2}{4 \hat{r}_{c}^8} .
\end{equation}
Higher order tidal interactions are considered in \cite{Damour:Nagar:2009,Vines:Flanagan:2010,Bini:Damour:Faye:2012}, but results therein can not immediately be included in the present comparison as they are not given in the form of Hamiltonians within the ADMTT gauge.

Finally, we solve the condition defining circular orbits
\begin{equation}
\frac{\partial H(\hat{r}_{c}, l_c)}{\partial \hat{r}_{c}} = 0 ,
\end{equation}
order-by-order in $\epsilon_{PN}$ for the ADMTT radial coordinate $\hat{r}_{c}$ and eliminate $\hat{r}_{c}$ from the Hamiltonian. We find that the result $H(l_c)$ agrees with $e(\hat{J})$ given by (\ref{ejPN}) for all the Hamiltonians shown in the present section (taking into account that $l_c = \hat{J} - q \hat{a}_2$).

\section{Conclusions}\label{conclusions_sec}

In this work we have investigated the influence of the internal structure of test bodies on their motion within the context of a multipolar approximation scheme. Corrections arising from the spin (dipole) and quadrupole moment were worked out explicitly for equatorial orbits, with aligned spin, in Kerr spacetime. In particular our explicit model for the quadrupole, which also allows for tidal deformations of the test body, goes beyond previous investigations in the literature in a multipolar context. 

Our comparison with recent numerical results for structureless bodies -- which take into account the self-force -- makes clear, that the corrections arising from the extendedness of the body can play a role in the description of the motion and should be carefully dealt with. A final statement regarding the magnitude of effects coming from different corrections (internal structure or self-force) depends on many factors. The figures in the present work immediately identify the relevant contributions needed to achieve a specific accuracy in the binding energy. As we have pointed out in the context of the binding energy, one should be careful when it comes to the parametrization of possible contributions in terms of different variables, cf.\ the corresponding discussion regarding parameters $l_c$ and $\hat{J}$. One should also stress at this point, that only the conservative parts of the self-force in \emph{Schwarzschild} spacetime -- which were the only ones readily available in the literature -- were taken into account in our comparison. 

Our work clearly indicates the necessity of future in-depth comparisons of different approximation schemes. In particular, as soon as results for the self-force in Kerr spacetime become available they should be compared to our findings here. Another interesting open question regards the possibility of extrapolating our results to less extreme mass ratios, i.e., intermediate mass ratios or even comparable masses, in the future. Whether such an extension of one of the existing multipolar approximation schemes can be consistently worked out is open for debate. A promising approach is to substitute masses or other parameters in a certain way, which at least for the self-force works astonishingly well \cite{LeTiec:etal:2011} and can also be understood as a change of expansion parameters \cite{LeTiec:Barausse:Buonanno:2012}. A comparison with post-Newtonian results (not expanded in the mass ratio) can serve as a guide to identify proper variable replacements. The post-Newtonian expansion of the gauge invariant relation $e(\hat{J})$ can further be used to match coefficients of an effective action with higher order spin couplings in the future. For example, if we would have kept the constant $C_{ES^2}$ for both objects in the post-Newtonian Hamiltonian $H_{\text{LO}}^{S^2}$ or $H_{\text{NLO}}^{S^2}$ (which can be derived directly from the effective action \cite{Porto:Rothstein:2008:2, Steinhoff:2011}), then a comparison with the result from the present paper would show that one has to set $C_{ES^2} = 1$ for the central black hole. The relation $e(\hat{J})$ is also very useful to check results of post-Newtonian or post-Minkowskian approximations.

The abovementioned extensions of the approximation method are required for the creation of gravitational wave template banks in the comparable mass case. Although numerical simulations are optimally suited for the late inspiral phase of such a binary, they currently are not able to cover the whole parameter space in an acceptable timeframe if both objects are spinning. A synergy of numeric and analytic methods is needed.

\begin{acknowledgments}
This work was supported by the Deutsche Forschungsgemeinschaft (DFG) through the grants STE 2017/1-1 (J.S.) and LA-905/8-1 (D.P.); and in an initial phase through the SFB/TR7 (J.S.), as well as the Japan Society for the Promotion of Science (JSPS) through the Global COE program G01 (D.P.). The authors thank the MPI for Gravitational Physics (Albert-Einstein-Institute) for the hospitality in 2011. Furthermore, D.P.\ thanks CENTRA for the hospitality in 2012. The authors are grateful to G.\ Sch\"afer, J.\ Hartung, and S.\ Hergt (University of Jena), as well as Y.N.\ Obukhov (Moscow State University) for helpful discussions. 
\end{acknowledgments}

\appendix

\section{Misprints within the effective potential} \label{sec_misprints}

There seem to be minor errors or misprints in the literature on the effective potential for the pole-dipole case \cite{Rasband:1973, Hojman:Hojman:1977, Tod:Felice:1976, Suzuki:Maeda:1998}. We will summarize here what misprints needs correction in order to achieve agreement with the result in the present paper, equations (\ref{pot1st})--(\ref{potlast}), and (\ref{U_definition}).

Equation \cite[(7a)]{Rasband:1973} must read
\begin{equation}
	\alpha = A \left[ \dots + \frac{2 s a}{r} \left( 3 + \frac{a^2}{r^2} \right) + \dots \right] ,
\end{equation}
where the dots are an abbreviation for correct terms in this section. In \cite{Hojman:Hojman:1977} there are two sign errors in the Appendix, namely
\begin{align}
	\gamma &= \dots - \delta^2 \Delta M^2 , \\
	k_i &= \dots - B_i \frac{J}{M r} + \dots .
\end{align}
In \cite[eq.\ (2.24)]{Suzuki:Maeda:1998} the expression for $Z$ must read
\begin{equation}
	Z = - \frac{\Delta \left( \frac{M S^2}{\mu^2 r^2} - r \right)^2 \mu^2}{\dots} .
\end{equation}
Finally, we find full agreement with \cite{Tod:Felice:1976} except for an overall sign of the spin $S$. We will argue in the next section how to correctly identify the co-rotating and counter-rotating cases.

\section{On the orientation of the spin}\label{sec_orientation}

Whether the spatial components of the spin vector $S^a$ allow a straightforward determination of the spin orientation depends on the sign choice in (\ref{S_vector}), which in turn depends on conventions for the spin tensor and the signature of spacetime.

The simplest way to identify the spin orientation for the sake of the present paper is via the angular momentum $J$ defined by (\ref{six_equations}),
\begin{equation}
J = - p_{\phi} + (g_{\phi t,r} p_{\phi} - g_{\phi\phi,r} p_t) \frac{S}{2 \underline{m} r} \,.
\end{equation}
In the weak field (large $r$) and slow motion limit we have $g_{\phi t,r} \approx 0$, $g_{\phi\phi,r} \approx - 2 r$, $p_{\phi} \approx - \underline{m} r^2 \dot{\phi}$, and $p_t \approx \underline{m}$. Therefore it holds
\begin{equation}
J \approx L + S \,,
\end{equation}
where the orbital angular momentum $L = \underline{m} r^2 \dot{\phi}$ is aligned to $\partial_z$ if $L>0$. This shows that for $S>0$ the spin is aligned to $\partial_z$.

More generally, one may define the angular momentum $J$ in the weak field and slow motion limit via the energy-momentum tensor as 
\begin{equation}
J^i \approx \epsilon_{ikl} \int d^3 x \, x^k T^{l0} ,
\end{equation}
which is still applicable if the spacetime possesses no rotational symmetry. From (\ref{em_tensor_singular}) we obtain
\begin{equation}
T^{l0} \approx m \dot{Y}^l \delta(x^i - Y^i) - \frac{1}{2} \partial_k ( S^{kl} \delta(x^i - Y^i)) ,
\end{equation}
and (\ref{S_vector}) leads to $S^i \approx \frac{1}{2} \epsilon_{ikl} S^{kl}$.
Therefore it holds
\begin{equation}
J^i \approx L^i + S^i,
\end{equation}
where $L^i := m \epsilon_{ikl} Y^k \dot{Y}^l$ is the usual Newtonian orbital angular momentum vector. Notice that for usual spherical coordinates $\partial_{\theta}$ points in the opposite direction as $\partial_z$ in the equatorial plane. Therefore $S^{\theta} < 0$ corresponds to a spin aligned to $\partial_z$.

\section{Existence of Equatorial and Circular Orbits}\label{sec_existence}

The conditions for equatorial orbits (\ref{eq_orbits_def}) and aligned spin (\ref{aligneds}) must be preserved under the time evolution, i.e.,
\begin{equation}
\dot{\theta} = u^{\theta} = 0 , \quad \dot{p}^{\theta} = 0 , \quad \dot{S}^{a\theta} = 0.
\end{equation}
Similarly, for the interpretation of the effective potential and the existence of circular orbits, it is important that under the condition $p^r = 0$ it follows that 
\begin{equation}
\dot{r} = u^r = 0 ,
\end{equation}
or in words, if $p^r = 0$ then there is either a turning point of the orbit or the orbit is circular (in both cases $u^r = 0$).

In order to prove these statements one can use the relation (\ref{mom_vel_relation}) and the equations of motion. It is further beneficial to use the coordinate time as the worldline parameter instead of the proper time, i.e., $u^t = 1$ (which is why we formulated our model (\ref{Jex1st}) in a reparametrization invariant manner). In the pole-dipole case the statements above where shown in, e.g., \cite{Rasband:1973, Hojman:Hojman:1977}. This calculation must be repeated for the quadrupole model (\ref{Jex1st}) now. As the details of such a calculation do not provide any physical insight, we simply relied on a brute force calculation using Mathematica \cite{Wolfram:2003} together with the free xTensor and xCoba packages \cite{MartinGarcia:2002,MartinGarcia:2008}. We found that the statements generalize to the quadrupole case.

\section{Quadrupole action}\label{sec_quadrupole_action}

In this section we show the extension of the point-mass Lagrangian that inspired the quadrupole model used in the present paper. It is actually more convenient to work with the Legendre transformation of the Lagrangian in the angular velocity, so we are technically considering a Routhian $R_M$ here. Combining models for quadrupole deformations induced by spin from \cite[eqs.\ (1) and (16)]{Porto:Rothstein:2008:2} and for \emph{adiabatic} tidal quadrupole deformations given by \cite[eq.\ (19)]{Goldberger:Rothstein:2006:2} or by \cite[eq.\ (5)]{Damour:Nagar:2009} leads to
\begin{eqnarray}
R_M &=& \mu \sqrt{u^2} - \frac{1}{\mu \sqrt{u^2}} B_{ab} S^a u_c S^{cb} - \frac{c_{ES^2}}{2 \sqrt{u^2}} E_{ab} S^a{}_{c} S^{cb} \nonumber \\
	&&- \frac{\mu_2}{4 (\sqrt{u^2})^3} E_{ab} E^{ab} - \frac{2 \sigma_2}{3 (\sqrt{u^2})^3} B_{ab} B^{ab} , \label{Routhian}
\end{eqnarray}
where $u^2 := u^a u_a$. Notice that some signs changed due to the adoption to our conventions. The parameters $\mu$, $c_{ES^2}$, $\mu_2$, and $\sigma_2$ are \emph{assumed} to be constant. One may write $\mu = m_0 + \frac{1}{2I} S^2 + \dots$, where $I$ can be interpreted as a moment of inertia and $m_0$ as an irreducible mass, see, e.g., \cite[eq.\  (3.28)]{Steinhoff:2011}.

The equations of motion for Lagrangians of the type used above have been worked out in a general fashion already in \cite{Bailey:Israel:1975} and were found to be of the form used in the present paper (see also \cite[sec.\ 5.2]{Steinhoff:2011}). In particular, \cite[eq.\ (19)]{Bailey:Israel:1975} provides a formula for the quadrupole $J^{abcd}$,
\begin{equation}
J^{abcd} = 6 \frac{\partial R_M}{\partial R_{abcd}}.
\end{equation}
It is straightforward to derive $J^{abcd}$ from this formula. But the equations of motion belonging to the $R_M$, shown in (\ref{Routhian}), preserve the spin supplementary condition $S^{ab} p_b = 0$ only approximately (in the sense of the multipole approximation introduced in sec.\ \ref{sec:quadmass}), while here we are enforcing this condition exactly. Therefore we made some minor changes in the result for $J^{abcd}$, which are, however, in accordance with the used approximation scheme. That is, we replaced $u^a$ by $p^a / \underline{m}$ and introduced the overall factor $m / \underline{m}$ in (\ref{Jex1st}) for the sake of reparametrization invariance (which was ensured by factors of $\sqrt{u^2}$ in the original expression).

\section{Explicit expressions for the binding energy}\label{sec_ca_results}

In this Appendix we provide explicit expressions for the expanded binding energy if $\hat{a}_1 = 0$. Although these can be easily derived from the more compact effective potential given in the main text using computer algebra, we display them here for the sake of completeness. The zeroth order solution to (\ref{circ_cond}) reads
\begin{equation}
\hat{r}_0 = \frac{l_c^2}{2} \left( 1 + \sqrt{1-\frac{12}{l_c^2}} \right) ,
\end{equation}
and the corresponding binding energy is given by
\begin{equation}
e_0 = \sqrt{\left( 1 - \frac{2}{\hat{r}_0} \right)\left( 1 + \frac{l_c^2}{\hat{r}_0^2} \right)} .
\end{equation}
It is straightforward to solve for higher orders in terms of $\hat{r}_0$ and $e_0$, resulting in
\begin{widetext}
\begin{eqnarray}
e(l_c) &=& e_0 + \frac{\hat{a}_2 l_c q}{\hat{r}_0 -2 } \hat{r}_0^{-3}(l_c^2 + \hat{r}_0^2)^{-1}\left[3 l_c^2 (\hat{r}_0-4) - 2\hat{r}_0^2\right]^{-1}  \bigg\{ l_c^4 (\hat{r}_0 -3)^2 (\hat{r}_0 -2) + \hat{r}_0^4 \left[6 + (2 + 2e_0 -3 ) \hat{r}_0 \right] \nonumber \\ 
&& - l_c^2 \hat{r}_0^2 (2 + e_0 \hat{r}_0)(6 + \hat{r}_0^2-6\hat{r}_0 )  \bigg\} \nonumber \\
&+& \frac{\hat{a}_2^2 q^2}{2}  (\hat{r}_0-2)^{-2} \hat{r}_0^{-3} (l_c^2 + \hat{r}_0^2)^{-2}( 2\hat{r}_0^2-3 l_c^2 (\hat{r}_0-4))^{-3} \bigg\{8 ( 4 + 4 e_0 -3) (-2 + \hat{r}_0)^2 \hat{r}_0^{10} \nonumber \\ 
&&+ 2 l_c^2 \hat{r}_0^8 \bigg[12 (\hat{r}_0-2)^2 [16 + (\hat{r}_0-9) \hat{r}_0]+ (1 + e_0) [\hat{r}_0 (449 + \hat{r}_0 ( 4 (31 - 3 \hat{r}_0) \hat{r}_0 - 399))-110]\bigg] \nonumber \\ 
&&- 3 l_c^{10} (\hat{r}_0-2) \bigg[2 (\hat{r}_0-3)^2 (\hat{r}_0-2 ) [24 + (\hat{r}_0-8) \hat{r}_0] - (1 + e_0) [\hat{r}_0 (3627 \nonumber \\ 
&&+ 2 \hat{r}_0 ( \hat{r}_0 (314 + 3 (\hat{r}_0-16) \hat{r}_0)-1056 ))-2484]\bigg] - l_c^4 \hat{r}_0^6 \bigg[6 (\hat{r}_0-4) (\hat{r}_0-2)^2 [116 + \hat{r}_0 (4 \hat{r}_0-49)] \nonumber \\ 
&& -(1 + e_0) [ \hat{r}_0 (6200 + \hat{r}_0 (-5601 + 2 \hat{r}_0 (1087 + 4 \hat{r}_0 ( 3 \hat{r}_0-47))))-2220]\bigg] \nonumber \\ 
&& - l_c^6 \hat{r}_0^4 \bigg[6 (\hat{r}_0-2)^2 [\hat{r}_0 (300 + \hat{r}_0 (-37 + (\hat{r}_0-6) \hat{r}_0))-504] - (1 + e_0) [10116  + \hat{r}_0 ( \hat{r}_0 (11801 \nonumber \\ 
&&+ \hat{r}_0 (\hat{r}_0 (1027 + 6 ( \hat{r}_0-21) \hat{r}_0)-4520))-17028)]\bigg] \nonumber \\ && - l_c^8 \hat{r}_0^2 \bigg[6 (\hat{r}_0-2)^2 [108 + \hat{r}_0 ( \hat{r}_0 (105 + 2 (\hat{r}_0-12) \hat{r}_0)-192)] - (1 + e_0) [26892 + \hat{r}_0 (\hat{r}_0 (41277 \nonumber \\ 
&& + \hat{r}_0 (\hat{r}_0 (4115 + \hat{r}_0 (27 \hat{r}_0-520))-17358))-52056)]\bigg] \bigg\} \nonumber \\ 
&+& \frac{1 + e_0}{\hat{r}_0-2 } \hat{r}_0^2 (l_c^2 + \hat{r}_0^2)^{-1} \left[ 2 \hat{r}_0^2 -3 l_c^2 (\hat{r}_0 -4 ) \right]^{-1} \bigg\{2 ( \hat{r}_0-1) \hat{r}_0^2 + l_c^2 \left[(16 - 3 \hat{r}_0) \hat{r}_0  -18 \right] \bigg\} \left( \frac{\underline{m}}{\mu} - 1 \right) \nonumber \\ 
&&+ (1 + e_0) \hat{r}_0^3 (l_c^2 + \hat{r}_0^2)^{-1} \left[l_c^2 (3 - \hat{r}_0) + \hat{r}_0^2\right] \left[3 l_c^2 (4 - \hat{r}_0) + 2 \hat{r}_0^2\right]^{-1} \frac{d \underline{m}/\mu}{d \hat{r}} + \Order(\epsilon^3).
\end{eqnarray}
\end{widetext}
where the $\underline{m} / \mu$ contributions are given by (\ref{mquad}) with $\hat{J} \approx l_c$, $\hat{r} \approx \hat{r}_0$, and $\hat{E} \approx 1+e_0$ inserted.

The self-force correction can be derived from the formulas in
\cite{LeTiec:Barausse:Buonanno:2012} and reads
\begin{equation}
\frac{e_{\text{TBB}}(l_c)-e_0(l_c)}{q} = -1 + \frac{1-\frac{5}{2} x_0}{\sqrt{1 - 3 x_0}} + \frac{1}{2} z_{\text{SF}}(x_0) ,
\end{equation}
where
\begin{equation}
6 x_0 = 1 - \sqrt{1-\frac{12}{l_c^2}} .
\end{equation}
The function $z_{\text{SF}}(x)$ is given in \cite{LeTiec:Barausse:Buonanno:2012} by a fit to a rational function with 5 parameters.

\section{Conventions \& Symbols}\label{dimension_acronyms_app}

In order to fix our notation, we provide some tables with definitions in this Appendix. The dimensions of the different quantities appearing throughout the work are displayed in table \ref{tab_dimensions} and table \ref{tab_dimensions_kerr}. Note that we set $c=1$, the dimension of the gravitational constant then becomes $[G]=m/kg$. Table \ref{tab_symbols} contains a list with the most important symbols used throughout the text. Latin indices denote 4-dimensional indices and run from $a = 0, \dots, 3$, the signature is (+,--,--,--). The Riemann tensor $R_{abd}{}^{c}$ is defined by
\begin{eqnarray}
\nabla_{[a} \nabla_{b]} a^{c} = \frac{1}{2} R_{abd}{}^{c} a^{d},
\end{eqnarray}
where $a^c$ is a generic vector. The volume form is given by $\eta_{abcd} = \sqrt{-g} \epsilon_{abcd}$, where $\epsilon_{abcd}$ is the completely antisymmetric Levi-Civita symbol with $\epsilon_{0123} = +1$. Notice that $\eta^{abcd} = \epsilon^{abcd} / \sqrt{-g}$ and $\epsilon^{0123} = -1$.

\begin{table}
\caption{\label{tab_dimensions}Dimensions of the quantities (general).}
\begin{ruledtabular}
\begin{tabular}{cl}
Dimension (SI)&Symbol\\
\hline
&\\
\hline
\multicolumn{2}{l}{{Geometrical quantities}}\\
\hline
1 & $g_{ab}$, $\sqrt{-g}$, $\delta^a_b$, $\eta^{a b c d}$, $\varepsilon^{a b c d}$, ${\xi^{a}}$\\
m& $s$, $Y^a$, $dx^a$ \\
m$^{-1}$& $\partial_a$, $\Gamma_{ab}{}^c$ \\
m$^{-2}$& $R_{abc}{}^d$, $E_{ab}$, $B_{ab}$ \\
&\\
\hline
\multicolumn{2}{l}{{Matter quantities}}\\
\hline
1& $u^a$, $U^a$, $\hat{p}^a$, $H$ \\
kg& $m$, $\underline{m}$, $p^a$, $\mu$, $R_M$ \\
kg\,m& $S^{ab}$, $S^{a}$, $S$ \\
kg\,m$^2$ & $J^{abcd}$, $Q^{abc}$, $Q^{ab}$\\ 
kg$/$m$^3$& $T^{\alpha \beta}$\\
&\\
\hline
\multicolumn{2}{l}{{Auxiliary quantities}}\\
\hline
1 & $\hat{p}^a$, $f^a$ \\
m$^{-4}$ & $\delta_{(4)}$\\
&\\
\hline
\multicolumn{2}{l}{{Operators}}\\
\hline
1 & $\rho^a_b$, $W^a_b$, $\hat{\rho}^a_b$, $X^a_b$   \\
m$^{-1}$& $\nabla_i$, $\frac{D}{ds} = $``$\dot{\phantom{a}}$'' \\
&\\
\end{tabular}
\end{ruledtabular}
\end{table}

\begin{table}
\caption{\label{tab_dimensions_kerr}Dimensions of the quantities (Kerr).}
\begin{ruledtabular}
\begin{tabular}{cl}
Dimension (SI)&Symbol\\
\hline
&\\
\hline
\multicolumn{2}{l}{{Geometrical quantities}}\\
\hline
1 & $\theta$, $\phi$\\
m& $t$, $r$, $M$, $a$\\
&\\
\hline
\multicolumn{2}{l}{{Matter quantities}}\\
\hline
kg & $E$, $p_t$, $p_r$, $S^{t\phi}$, $S^{r\phi}$, $U_\pm$\\
kg\,m & $J$, $p_\phi$, $p_\theta$, $S^{tr}$, $S_{\theta}$\\
&\\
\hline
\multicolumn{2}{l}{{Auxiliary quantities}}\\
\hline
1 &  $\alpha$, $\beta$, $\gamma$, $\delta$, $\sigma$, $e$, $C_{ES^2}$, $\hat{R}$, $\hat{U}$\\
  &  $\hat{r}$, $\hat{J}$, $\hat{E}$, $\hat{a}_1$, $\hat{a}_2$, $q$, $j_2$, $k_2$, $l_c$\\
m &  $\rho$\\
m$^2$ & $\Delta$\\
kg$^{-2}$\,m$^{-5}$ & $A_0$\\
kg$^2$\,m$^{(n)}$& $A_{1}^{(n=3)}$, $A_2^{(4)}$, $A_3^{(5)}$, $A_4^{(7)}$\\ 
kg$^{-1}$ & $c_{ES^2}$\\
kg\,m$^{4}$ & $\mu_2$, $\sigma_2$\\
&\\
\end{tabular}
\end{ruledtabular}
\end{table}

\begin{table}
\caption{\label{tab_symbols}Directory of symbols.}
\begin{ruledtabular}
\begin{tabular}{ll}
Symbol & Explanation\\
\hline
&\\
\hline
\multicolumn{2}{l}{{Geometrical quantities}}\\
\hline
$g_{a b}$ & Metric\\
$\sqrt{-g}$ & Determinant of the metric \\
$\delta^a_b$ & Kronecker symbol \\
${\xi^a}$ & Killing vector\\
$x^{a}$, $s$ & Coordinates, proper time \\
$Y^a$ & Worldine\\ 
$\Gamma_{a b}{}^c$ & Connection \\
$R_{a b c}{}^d$& Curvature \\
$E_{ab}$, $B_{ab}$ & Curvature (electric, magnetic)\\
$M$, $a$ & Kerr (mass, parameter) \\  
&\\
\hline
\multicolumn{2}{l}{{Matter quantities}}\\
\hline
$u^a$ & Velocity \\
$m$, $\underline{m}$ & Mass (Frenkel, Tulczyjew)\\
$\mu$ & Mass-like parameter \\
$p^a$ & Generalized momentum \\
$S^{ab}$, $S^a$, $S$ & Spin (tensor, vector, length) \\
$T^{a b}$ & Energy-momentum tensor\\
$R_M$, $H$ & Routhian, Hamiltonian \\ 
&\\
\hline
\multicolumn{2}{l}{{Auxiliary quantities}}\\
\hline
$\mu_2$, $\sigma_2$, $c_{ES^2}$, $C_{ES^2}$, $j_2$, $k_2$ & Coupling constants \\
$\hat{R}$ & Dimensionless radius\\
$q$ & Mass ratio\\
$U_\pm$, $\hat{U}_\pm$& Effective potential\\
&\\
\hline
\multicolumn{2}{l}{{Operators}}\\
\hline
$\rho^a_b$, $X^a_b$, $W^a_b$ & Spatial projectors \\
$\eta^{abcd}$, $\varepsilon^{abcd}$ & Permutation symbols\\
$\partial_i$, $\nabla_i$ & (Partial, covariant) derivative \\ 
$\frac{D}{ds} = $``$\dot{\phantom{a}}$'' & Total derivative \\
&\\
\end{tabular}
\end{ruledtabular}
\end{table}

\bibliographystyle{unsrtnat}
\bibliography{consquantquad}

\end{document}